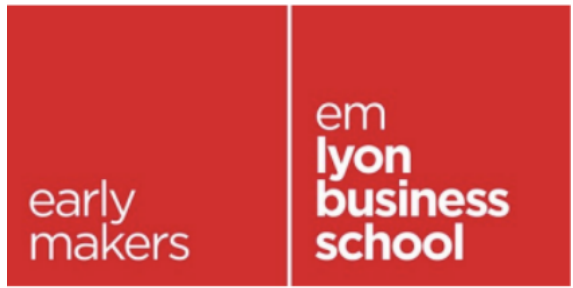

# Empirical Study on the Factors Influencing Stock Market Volatility in China

By

Jingchu Zhang

Emlyon Business School

November 2024

# Abstract

This paper mainly utilizes the ARDL model and principal component analysis to investigate the relationship between the volatility of China's Shanghai Composite Index returns and the variables of exchange rate and domestic and foreign bond yields in an internationally integrated stock market. This paper uses a daily data set for the period from July 1, 2010 to April 30, 2024, in which the dependent variable is the Shanghai Composite Index return, and the main independent variables are the spot exchange rate of the RMB against the US dollar, the 10-year treasury bond yields in China and the United States and their lagged variables, with the effect of the time factor added. Firstly, the development of the stock, foreign exchange and bond markets and the basic theories are reviewed, and then each variable is analyzed by descriptive statistics, the correlation between the independent variables and the dependent variable is expanded theoretically, and the corresponding empirical analyses are briefly introduced, and then the empirical analyses and modeling of the relationship between the independent variables and the dependent variable are carried out on the basis of the theoretical foundations mentioned above with the support of the daily data, and the model conclusions are analyzed economically through a large number of tests, then the model conclusions are analyzed economically. economic analysis of the model conclusions, and finally, the author proposes three suggestions to enhance the stability and return of the Chinese stock market, respectively.

# List of Figures

# List of Tables

# Table of Contents

# 1 Introduction

## 1.1 Background of the research topic

In 1984, Shanghai Feile Audio Co., Ltd. issued shares to the whole society for financing, and this stock, which could be traded in the Jing'an Securities Department, opened the door of the new Chinese stock market and became the first stock in the real sense. After that, China's stock market developed rapidly, from 8 stocks listed on the SSE when it was established in 1990 to more than 4,500 companies listed on the SSE and SZSE by the end of 2022, with a total market capitalization of 80 trillion yuan. During this period, China's stock market has experienced the outbreak of the stock market in the 1990s, the reform of the stock split in 2005, the financial crisis and the "four trillion" rescue policy in 2008, the stock market crash in 15 years, and the impact of many events in recent years, such as trade wars, the Corona epidemics and regional conflicts, and stock market volatility is large, especially in the crisis in 2008 and the stock market crash in 15 years, the volatility of China's stock market has reached a high level. The volatility of China's stock market during the crisis in 2008 and the stock market crash in 2015 was even two to four times higher than that of other normal years. Although stock market volatility is a normal performance of the market, which is conducive to the flow of social funds from the surplus side to the underfunded side, and the stock market without volatility will lose its function of resource allocation and the significance of its existence, excessive volatility of the stock market will cause great damage to the operation of China's economy and investors in the market, and therefore scholars continue to study the factors of volatility of China's stock market and make suggestions for the development of China's stock market and how to avoid the risk of investors. Therefore, scholars continue to study the factors of China's stock market volatility and make suggestions for the development of China's stock market and how investors can avoid risks.

Nowadays, economic globalization has reached an unprecedented height, China's door to the world is more and more open, the international capital, finance and trade links are becoming more and more close, so it can be said that any wind blowing in the international financial market always pulls every move of China's capital markets. On November 30, 2015, the RMB successfully joined the "basket of currencies" (SDR), marking a milestone in its internationalization. Secondly, the expansion of RMB global clearing centers and the use of RMB as the settlement currency for oil trade between China and Russia have made us hail the arrival of the oil RMB era. In addition, China is vigorously developing the "Belt and Road" policy and the Asian Infrastructure Investment Bank (AIIB), which will open up domestic and international transportation routes and promote global exchanges. Finally, the Shanghai

Composite Index is trying to be included in the MSCI index of emerging markets; the central bank has eased the conditions for international lobbying to invest in China's stock market and inter-bank bond market; and the Fed has raised interest rates, which has a serious impact on China's economy. The pulse of the Federal Reserve's interest rate hike is seriously affecting the beating of the Chinese stock market; and Chinese capital is vigorously merging and acquiring foreign enterprises, accelerating the pace of going out, etc., all of which are evidence of the deepening of global economic, financial and trade ties.

1.2 Purpose and significance of the research topic

The research in this paper has two significances: one is the theoretical significance of the research and the other is the practical significance of the research. First of all, from the theoretical significance, most of the existing studies on the factors of China's stock market volatility consider the role of the bond market or the exchange market on the stock market volatility separately, and seldom combine the two influencing factors, or combine the two but ignore the impact of macroeconomic variables on stock market volatility. In addition, different scholars often get opposite conclusions when using different models to study the impact of certain factors on stock market volatility. On the basis of previous scholars, this paper comprehensively researches the role of U.S. Treasury rates, RMB-dollar exchange rates and macroeconomic variables on stock market volatility and analyzes what kind of factors will play a role in stock market volatility and in what direction by taking advantage of the existing unique time to obtain a larger amount of data, which will form a certain complementary role to the results of previous research, and is of great theoretical significance.

Secondly, from a practical point of view, stock market volatility is the basic characteristics of the stock market, the allocation of social resources play an important role, but the stock market fluctuations will disrupt the process of resource allocation, thereby affecting the operation of the macro-economy, in addition to the majority of small and medium-sized investors will be in the large stock market fluctuations in the losses suffered. Therefore, this paper studies the impact of bond market, exchange market and macroeconomic volatility on the stock market, can give managers to formulate fiscal and monetary policy, investors to avoid investment risks, etc., so the study of China's stock market volatility of the impact of factors of great practical significance.

Due to the increasingly close economic, trade and financial ties among countries and regions around the world in recent years, the international financial market has been characterized by the gradual deregulation of financial control and the deepening of financial innovation, and Financial Market Integration has become an irreversible development trend. The so-called Financial Market Integration refers to the process and trend of the financial markets of the world's countries and regions becoming increasingly interconnected, deepening their mutual influence, and gradually forming a unified financial market under the impetus of financial liberalization, financial innovation, and advances in information technology. During the period of June-August 2015, China's capital market ushered in the stock market bloodletting, at the same time, the RMB exchange rate was fragile and unbearable, the capital flight situation was disastrous, and the situation of "stock and foreign exchange double kill" was staged with extraordinary ferocity, which showed that the influence of international factors on China's stock market was not comparable to that of history. As one of the important forces in the world and the representative of emerging economies, China must carefully measure the development trend and direction of the global economy, in order to maintain the power of not being stranded in the surging tide of the world economy. In this context, how to accurately describe the interdependence of financial markets has become an important topic for scholars at home and abroad. Exploring the interdependence of international financial markets is of great practical significance for understanding and grasping the laws of economic and financial development, improving the accuracy of financial decision-making by investors and supervisory authorities, and reducing the risk of decision-making, which is not only an important premise for investors to diversify their investments to effectively avoid risks, but also an important policy significance for the supervisory authorities to avoid the volatility of the domestic market and the risk of instability in the financial market caused by the external market. Policy significance. It is because and the international environment on China's capital market is increasingly profound, so this paper focuses on the study of domestic and foreign market factors on China's A-share market influence degree. Secondly, the author believes that this research topic is more forward-looking, although China's capital market is not yet fully open to the outside world, foreign capital into the market channels are still subject to squeeze, but with the development of the foreign economy and the unswerving pace of reform and opening up, China's road to a mature capital market more and more flat, it can be expected that one day in the future, China's stock market will be completely open to the outside world, to set up a capital market system like the United States, as mature as the perfect, and also like a magnetic field, the Chinese stock market will be fully open to the outside world. It can be expected that one day in the future, China's stock market will be completely open to the outside world and establish a mature and perfect capital market system like the United States, and like a magnetic field to attract international lobbying investment. Although the current Chinese market has already been "water" enough, but the input of foreign capital is bound to be for China's economic development will continue to inject a steady stream of living water, so that capital can better irrigate the real economy.

When discussing the influence mechanism of China's stock market, this paper selects the United States as an example for research, mainly based on the following considerations. First of all, as the world's largest economy, the U.S. financial market has an important leading role in the global context. In particular, the U.S. Treasury rate is often regarded as the benchmark of the global bond market, and its fluctuations directly affect the global capital flow and market risk appetite. Therefore, using U.S. Treasury rates as a research variable helps to assess its potential impact on the stock markets of emerging markets such as China. Second, as the world's top two economies, China and the U.S. have close and complex interactions in the economic and financial fields. The fluctuations of the U.S.-China exchange rate, a key variable in measuring the interaction between the two economies, have a significant impact on the Chinese stock market. Changes in the exchange rate not only affect the profit expectations of multinational corporations, but also change the cross-border volatility of capital, which in turn has an impact on the stock market. Finally, U.S. economic policies, especially monetary policy adjustments, not only affect investment and consumption within the U.S., but also affect other countries through various channels. By studying the mechanism of the U.S. Treasury rate and the U.S.-China exchange rate on China's stock market, it is possible to gain a deeper understanding of how international policy changes affect the real economy through the financial market.

### 1.3 Research content and approach

The article chooses the 10-year government bond yield represented by the United States as a proxy variable for international factors, and at the same time chooses the spot exchange rate of the RMB against the US dollar, the 10-year government bond yield and the daily closing price of the Shanghai Composite Index (SSE) as the proxies for the foreign exchange market, the bond market and the stock market respectively, and studies in depth the influence of the exchange rate and the domestic and foreign bond yields on the SSEI of China in order to reflect the interrelationships among the variables with the help of a reasonable econometric model. In order to reflect the interrelationships among the variables through a reasonable econometric model.

In order to comprehensively measure the interconnection mechanism between foreign factors and domestic securities market, stock market, exchange rate market and bond market, to reflect the influencing factors of China's A-share market more comprehensively, and then to provide guidance for stabilizing China's stock market, the author does not limit the research perspective to the influence of domestic factors on the stock market, but extends it to the joint influence of domestic and foreign factors, and tries to stand in the perspective of economic globalization, to derive an appropriate form of modeling to provide good decisions and suggestions for the long-term sound development of China's stock market. Specifically, it is hoped that through the modeling research in this thesis, we can find out the direction of the influence of individual variables on the volatility of stock index returns and the optimal lag period. For example, if the independent variable, the U.S. 10-year Treasury rate, is a variable that can have a significant impact on the volatility of the Shanghai Composite Index, it means that the relevant regulators of China's securities market need to pay close attention to the daily dynamics of the U.S. Treasury market.

This paper utilizes the combination of theory and empirical evidence, qualitative and quantitative, text and charts. Theoretically, the flow-oriented theory and IS-LM-IA theory between the exchange market and the stock market; the discounted cash flow theory and the asset pricing theory between the bond market and the stock market are introduced respectively, and the capital asset pricing theory is verified and extended according to the actual situation.

Empirically, GARCH is mainly used to simulate the stock market volatility, and then the vector autoregressive model (VAR) is used. The monthly data from July 2010 to April 2024 are used as the samples, and the daily stock market volatility is firstly constructed with the GARCH model, and then the U.S. treasury bond yields, the Chinese government bond yield and the USD-RMB exchange rate are tested for stationarity. After that, the autoregressive distributed lag (ARDL) model and lagged nonlinear model, which are a combination of multivariate regression analysis and vector autoregressive model (VAR), are adopted to fully take into account the impact of the lag period of the variables, which is very suitable for the Chinese capital market with time lag effect, and reflect the essence of the problem more comprehensively and accurately, which is even more appropriate for the study of the non-stationary time series data.

In the validation part of the modelling results, many statistical tests are carried out using econometrics software STATA, including unit root test, cointegration test, heteroskedasticity test and Granger causality test on the stationarity of the time series data in order to further ensure the accuracy of the model, and the model finally passes all the tests. And relevant suggestions and countermeasures are proposed at the end of the article to escort the long-term stationary operation of the Chinese stock market.

1.4 Innovations in this paper

At present, domestic and foreign research on the correlation between China's stock market and bond market is relatively comprehensive, and there is a relative lack of research on the stock market and foreign exchange market, or research on the correlation between the foreign exchange market and the stock market, the bond market and the stock market respectively. Based on the innovation of the times, the author believes that the development of China's stock market not only has a coupled relationship with the domestic capital market, and the link with foreign markets is also deepening, the correlation analysis of the domestic stock market and bond market will lead to research horizons are blocked. Although in the past, when studying the impact of interest rates on stock prices alone, the exchange rate factor may be used as a control variable before modeling. However, the focus of the research is still limited to the interest rate as the main independent variable, including the summary of references, the support of theoretical foundations, the formulation of research hypotheses and so on, while the relationship between the control variable and the dependent variable is not studied in depth. Therefore, the article comprehensively combines the impact of China's foreign exchange market and bond market on the stock market, and at the same time, comprehensively considers the impact of foreign markets on China's stock market, both in terms of content and empirical evidence are more comprehensive, more practical value.

First of all, compared with the research on the relationship between exchange rate and stock price or interest rate and stock price alone, the article analyzes the interest rate and exchange rate factors together, which is still lacking in domestic and foreign academic circles. Moreover, most of the previous studies are directly regressed on the exchange rate and interest rate as the independent variables, without considering the correlation between the interest rate and the exchange rate, which will lead to serious multicollinearity problems in the model. Based on this, the author conducted a correlation test between the exchange rate and interest rate before modeling and chose the method of principal component regression to exclude the multicollinearity, which makes the research conclusions more reliable. In addition, most of the existing studies use the VAR model of multiple linear regression, while there are fewer analyses using the ARDL model.

Secondly, when studying the impact of bond market on stock market, the Chinese and U.S. treasury indices were chosen as the research objects respectively. Most of the previous studies have broken down the domestic bond market and analyzed the impact of the exchange bond market and the interbank bond market on the stock market separately. The author believes that choosing both domestic and international factors for measurement would be more meaningful for the study and better reflect the interconnections between the markets. Furthermore, the study period covers the last three years, and the sample data are new and authoritative.

## 2 Literature review and analysis of the current market situation

### 2.1 Literature review

Since the stock market, the exchange market and the bond market are three important sub-markets in China's capital market, scholars at home and abroad have done more research on them, but basically, they have analysed the relationship between them. The following is the literature review to the research methods and results of domestic and foreign scholars.

#### 2.1.1 Literature review on exchange rate and stock index linkages

In terms of foreign studies, Aggarwal (1981), by studying the U.S. stock market and exchange rate market, found that there is a tendency of isotropic change between the exchange rate and the stock price, i.e., when the exchange rate rises, the stock price rises in tandem, and when the exchange rate declines, the stock price declines in tandem. However, Soenen and Hennigan (1988), who also studied the U.S. stock price and exchange rate, verified a negative correlation between the two. In addition, Bahmani Oskooee (1992) conducted cointegration test and Granger causality test on the US S&P index and exchange rate, and the empirical results showed that the cointegration test between the two could not be passed, i.e., there was no long-run equilibrium relationship, but the results of Granger causality test were bi-directional causality. Yu, Q (1996), also applying the method of Granger causality test, found that There is a unidirectional causality from the exchange rate to the stock price in the Hong Kong, China market, i.e., the exchange rate is the Granger cause of the stock price, but the stock price is not the Granger cause of the exchange rate.

Pan (2001) studied that the exchange rate had a significant effect on stock prices during the Asian financial crisis. Harald Hau and Helene Rey (2006) developed an equilibrium model of exchange rate, stock prices and capital flows under incomplete foreign exchange risk trading conditions, and found that when the exchange rate rises, i.e., when there is a depreciation of the domestic currency, the price of the domestic stock market also rises, thus obtaining a higher investment return in the domestic capital market relative to other countries. It is found that when the exchange rate rises, i.e., when the national currency depreciates, the price of the domestic stock market will rise, and the domestic capital market will have a higher investment return relative to other countries, thus verifying the positive correlation between the exchange rate and the stock price. Ming-Shiun Pan, Robert Chi-Wing Fok, and Y. Angela Liu (2007) conducted a study on the relationship between the exchange rate and the stock price before and after the financial crisis, and selected seven countries (regions) represented by Japan, and the exchange rate and the stock price in Japan. The study finds that before the financial crisis, Japan's exchange rate and stock prices were not as high as they were before the financial crisis. It is found that before the financial crisis, there is a significant effect of exchange rate on stock price in Japan, Hong Kong, and Malaysia, while after the financial crisis, the effect of exchange rate on stock price no longer exists in Malaysia. It can be seen that the effect of exchange rate on stock price varies between different capital market periods and different countries (regions). Leo Michelis and Cathy Ning (2010), on the other hand, analyzed the dependence of Canadian stock market returns on the Canadian dollar-US dollar exchange rate, and the authors, by producing a joint contour plot of stock market returns and foreign exchange market returns, found that the contours are steeper in the negative third quadrant than in the positive first quadrant. Therefore, the authors conclude that the stock market and the exchange rate show an asymmetric correlation, i.e., the correlation is more significant when the yields between the two markets are both negative.

K. Kennedy and F. Nourizad (2016) examines how fluctuations in the exchange rate between the U.S. dollar and the euro influence the volatility of the U.S. stock market. Utilizing a GARCH (1,1) model and analyzing weekly data from January 1, 1999, to January 25, 2010, the authors assess the impact of exchange rate volatility on stock returns, while accounting for other factors that may drive stock return volatility. Their findings indicate that increased exchange rate volatility has a positive and statistically significant effect on the volatility of stock returns, even when controlling for major drivers of financial volatility.

In domestic research, Zhu Xinrong and Zhu Zhenyuan (2008) established a GARCH model based on daily data from 2005 to 2007 by taking the logarithm of the median price of the exchange rate of the RMB against the US dollar as a proxy for the exchange rate factor, and at the same time taking the logarithmic form of the closing price of the Shanghai Composite Index as a proxy for the stock price, and then established a GARCH model through the unit root test, the cointegration test, the causality test, the ARCH test, and the GARCH test. GARCH test, finally, in the regression equation with stock price as dependent variable and exchange rate as independent variable, the optimal lag between stock price and exchange rate is determined to be 1 period and there is a negative correlation between them. The authors concluded that the exchange rate has an impact on stock price mainly through two ways: trade balance and capital flow. When the nominal exchange rate rises, it indicates a depreciation of the RMB, at which time the payoffs of holding RMB assets are relatively lower, thus leading to more international capital investing in foreign currency assets and withdrawing from the Chinese stock market, which in turn leads to a downward movement in stock prices and a decline in stock market returns.

Song Qin and Song Rong (2009) conducted a study on the transmission mechanism of exchange rate on stock prices in China. The authors categorize the transmission of exchange rate on stock prices into two ways: physical transmission and financial transmission, under which can be specifically subdivided into a variety of transmission effects. Among them, the financial structure effect under the financial transmission mechanism argues that if a country's exchange rate rises and the local currency depreciates, the loopholes in the financial structure will make the country's capital outflow phenomenon aggravate, which will lead to a decrease in the domestic demand for purchasing stocks, and a decrease in stock prices. Therefore, there is a negative correlation between the exchange rate and stock prices under the financial structure effect. Li Zhong (2009) found that changes in the exchange rate can cause changes in stock prices and affect the significant factors of stock prices. Bi-Qiong Zhang and Yue Li (2002) found a long-run cointegration relationship between the RMB exchange rate and both Shanghai and Shenzhen markets and Hong Kong's Hang Seng Index through ARDL model. Ma Shuang, Zhang Xiaolin, and Zhang Xiaoqi (2010) empirically analyzed the relationship between the exchange rate of RMB against US dollar and China's Shanghai Composite Index with the sample of data from 2005-2010. The cointegration test shows that there is a long-run equilibrium relationship between the exchange rate and the stock price, and a negative correlation between the two in the short run is derived from the error correction model. The rise in exchange rate makes the RMB depreciate, which in turn leads to an increase in exports, a decrease in imports, a strengthening of the domestic economy and an increase in GDP, which ultimately leads to an increase in stock prices and an increase in stock market returns.

Zhu Mengnan and Liu Lin (2010) studied the interrelationship between short-term international capital flows, exchange rates and asset prices in the capital market. The authors concluded that in a relatively short period of time, the inflow of international capital will lead to the appreciation of the RMB and an increase in stock prices, while the rise in exchange rates and the rebound in the stock market will further attract international lobbying capital to enter the Chinese stock market, thus forming a virtuous circle, with strong positive correlation between the three. Zhang Bing, Feng Sixian, Li Xindan et al. (2008) found that the exchange rate significantly affects stock price changes in the long run, while the short-term exchange rate changes have a time lag effect on stock prices. Lin Yiren (2011) studied the relationship between the Shanghai Composite A-shares, B-shares and the median exchange rate of RMB against the US dollar, and established a regression model by logarithmizing the daily data. The empirical results show that exchange rate fluctuations can have a more significant impact on China's A-shares and B-shares in the Shanghai Composite Index, and also illustrate the long-run equilibrium relationship between the two through the cointegration test. Yu Chao (2013) studied the correlation between the foreign exchange market and China's capital market from the perspective of exchange rate reform, interest rate marketization and the opening of the capital account through the test of least squares and VAR model. The results show that with the increase of exchange rate marketization and the deepening of capital market opening, the exchange rate mainly affects the stock price through the money supply channel.

Xie, Xiaowen, Fang, Yi, and Liang, Lulu (2013) investigated the nonlinear relationship between exchange rate and stock price by nonlinear Granger causality test using exchange rate, CSI 100 index and CSI 500 index as sample data from 2005 to 2012. The authors use the highest and lowest points of stock prices as the boundary, and divide the study period into bull market, bear market, rally and shock intervals. By building a VAR model to eliminate the linear part, and the BDS test on the residual term, it is found that the nonlinear relationship between the exchange rate and the stock index does exist, and the linear relationship is different in different stock market intervals. In particular, during the rebound period, there is no nonlinear causality between the exchange rate and the CSI 100 index, while it exists with the CSI 500 index. During the bull market, there is a nonlinear causal relationship between the exchange rate and both the CSI 100 and CSI 500 indexes. As can be seen, the non-linear relationship between the two varies according to the stock market cycle.

### 2.1.2  Literature review on interest rate and stock index linkages

Interest rate is the representative indicator of bond market and stock price is the main research object of stock market, so the research based on the relationship between interest rate and stock price is very closely connected with the correlation research of bond market and stock market. Foreign scholars Chordia, Sarkar and Subrahmanyam (2001) found that the inverse pattern of stock market and bond market is valid, and John T. Scruggs and Paskalis Glabadanidis (2003) used the capital asset pricing (CAPM) model as a benchmark, considered the effect of inter-period variation factors, and used the great likelihood estimation method to estimate the interest rate and the bond price, and the interest rate and the stock price are the main indicators of the stock market. , applying great likelihood estimation, extend the model to a two-factor model of risk pricing, using the growth rate of stock prices as the stock market return, and the bond market return to capture the correlation between the stock market and the bond market. The authors build a conditional second-order matrix using monthly data from the late World War II period, and the empirical results find that the correlation between the two markets varies with the span of the sample interval. The correlation coefficients between the two markets are negative in the late 1950s and early 1960s and positive in the mid-1960s and beyond. Therefore, the authors believe that the economic cycle affects the relationship between the stock and bond markets, and this empirical result is a novelty of the article. Earlier studies have suggested that there is a constant correlation between the two and it does not change over time. In addition, negative returns on the bond market of the same size are associated with a larger shock to the volatility of the conditional variance of the stock market than positive returns, i.e., the bond market exhibits an asymmetric response to the stock market.

Fleming, Kirby and Ostdiek (2003), on the other hand, examine the impact of time variation on return spillovers between the stock and bond markets by analyzing daily stock yields and Treasury yields. First, changes in real interest rates under different time periods make the positive correlation between stock and bond markets more significant. Second, consistent market expectations of future returns also cause their positive correlations to hold. However, inflation expectations over time make the negative correlation increasingly pronounced. In addition, cross-market hedging and investors' risk-averse buying sentiment is another reason for the negative correlation to hold. In terms of uncertainty in the market environment, the authors measured various aspects such as the expected future volatility of stocks and the uncertainty of economic conditions, and found that the higher the uncertainty faced by the stock market, the more pronounced the negative correlation between the stock and bond markets.

Moreover, this view is supported by Robert Connolly, Chris Stivers and Licheng Sun (2005) who examined the uncertainty in the market environment and the relationship between the stock and bond markets. The authors concluded that the moderate positive correlation between stock and bond markets holds in the long run, but the relationship may also be negative in the short run, and the correlation between stock prices and interest rates is not uniquely determined.

In terms of domestic studies, Zhang Shaobin and Qi Zhongying (2003) used interest rates and stock price indices to build a regression model, and the empirical results showed that there is a negative correlation between interest rates and stock prices. Yang Huimin (2009) established a VAR model using the Shanghai Composite Index as the dependent variable and GDP, money supply and interest rate as the independent variables, and the results of the study showed that there was also a negative correlation between interest rates and the Shanghai Composite Index. Chang Lili (2009) argued that the long-term equilibrium relationship between China's bond market and stock market does not exist, on the contrary, the independence is more significant, and the relationship is largely subject to the government's policy behavior. Yang Jie (2010) studied the relationship between interest rates and stock prices in China based on the theory of money demand with quarterly data from 1994 to 2008. Through the establishment of a VAR model, cointegration test, impulse response analysis and variance decomposition, Yang Jie (2010) argued that there is a long-run equilibrium relationship between interest rates and stock prices, and that given a positive shock to the interest rate, the stock price will react by falling, and this reaction process will last for a long time. And this reaction process will last for a long time. Therefore, there is a negative correlation between interest rates and stock prices.

Yuan Chen and Fu Qiang (2010) conducted an empirical analysis of the relationship between stocks, bonds and other financial submarkets. The authors covered the research period from 2007 to 2010, and classified the stock prices into the uptrend, crisis and post-crisis periods according to the historical performance of the stock prices, and introduced dummy variables into the model, and through the ADF test and the Q-statistics test, they found that the relationship between the markets in different periods also showed different characteristics. When the stock price is in the downward channel, the substitution effect between the stock market and the bond market is obvious, and the bond market yield rises; and when the stock price is in the upward channel, it will also drive the bond market to be strong together, and the bond market yield also goes up. It can be seen that the correlation between stock prices and interest rates may be both positively and negatively correlated, and the bond market will become a position for investors to transfer risks during the bear market, while also obtaining excess returns during the bull market, so China's financial regulators should pay full attention to the good development and sound operation of the bond market, so as to make it become a powerful driver for the improvement of the capital market.

Sun Qianyou (2013) measured the linkage between the stock market and bond market from the dimensions of interest rate, money supply, inflation, exchange rate, fixed asset investment and consumption, etc. Through the empirical test of VAR model and VECM model, it was found that there is indeed a weak correlation between interest rate and stock price. Liu Jianchun (2005), on the other hand, found a positive correlation between stocks and bonds. Zheng, Zhenlong, and Chen, Zhiying (2011), using the Shanghai Composite Index and CITIC All Bond Index yields, found that the correlation between the stock market and the bond market showed different relationships over different study periods. Zeng, Zhijian, and Jiang Chuan (2007) proved the existence of a long-run equilibrium relationship between the stock market and bond market yields through an empirical study. In addition, Wang Yuan and Li Fan (2014) investigated the correlation between stock and bond markets in China, the United States and between the two countries, respectively. On the theoretical side, asset pricing model and portfolio theory are mainly used to develop the discussion, and it is believed that interest rates are negatively correlated with the stock market. The empirical results show that the see-saw effect between the stock and bond markets holds in both China and the United States. In addition, the authors further explore the interaction between the capital markets of China and the United States, and find that there is a positive correlation between the Chinese stock market and the U.S. bond market in the long run, i.e., when the Chinese stock price goes up, the U.S. bond market also goes up, and the bond market yield increases.

Pan Weihong (2015) explored the relationship between U.S. long-term Treasury bond prices and China's SSE index by empirically analyzing daily data between 2005 and 2012. The study used the unit root test to verify the stationarity of the time series, the cointegration test for analyzing the long-run equilibrium relationship, and the Granger causality test to identify the direction of causality. It is found that the correlation between the two in different economic stages is significantly different: in the pre-bull market, the U.S. Treasury prices are negatively correlated with the SSE index, while in the period of global economic recovery, the two are positively correlated; however, there is no cointegration between the two during the bull market and the financial crisis. The paper points out that the U.S. Treasury bond price trend can provide a reference for the prediction of China's stock market under specific conditions, but its prediction effect is affected by a variety of complex factors, and its practical application value is limited.

Xueying Zhang, Chao Feng, and Shiqun Ma (2023) deeply discuss the spillover effect between China and U.S. Treasury bond yields and its influencing factors. The study employs a dynamic Nelson-Siegel model to extract the level, slope and curvature factors of the Chinese and U.S. Treasury yields, which comprehensively reflect the characteristics of the term structure of interest rates. Subsequently, the time-varying parameter vector autoregression (TVP-VAR) model is applied to analyze the spillover effects of the Chinese and U.S. bond yield factors and their influencing factors. The results show that the spillover effect between Chinese and U.S. bond yields is highly time-varying and asymmetric, and that the depreciation of the RMB exchange rate and China's short-term cross-border capital flows have a significant impact on the spillover effect.

### 2.1.3 Literature review on the relationship between exchange rates, interest rates and stock prices

Compared with the separate correlation studies between exchange rates and stock prices and interest rates and stock prices, the simultaneous analysis of the impact of exchange rates and interest rates on stock prices is much less common in both domestic and international academic circles.

Michael J. Brennan and Yihong Xia (2006) studied the relationship between risk premiums in the foreign exchange market and the corresponding capital market, and the authors concluded that when there is no speculative arbitrage in the international market, exchange rates, interest rates and stock market yields of various countries will show more stable correlations. Jin Tao (2010) used structured vector autoregressive model, SVAR model and ITH to study the lagged correlation and spillover effect among exchange rate, interest rate and stock price, and the empirical results show that there is basically no correlation among the means of the three, but the correlation among the variances is more significant. Wen Yuangen (2011) took the RMB exchange rate, one-year deposit rate and Shanghai Composite Index from 2007 to 2009 as the research samples to study the relationship among the three during the financial crisis, and established a regression model to conduct Granger causality test, impulse response analysis and variance decomposition. The results of the empirical analysis show that exchange rate changes can indeed have an impact on stock prices, but the extent of the impact of interest rate changes on stock prices is not significant. Li Cheng, Guo Zheyu, Wang Ruijun (2014) used VAR-GARCH-BEKK model for empirical analysis, the results show that the exchange rate, interest rates on the stock market have a spillover effect, and in the long run, the exchange rate than the interest rate of the spillover effect is more durable. Yan Hong (2015) first studied the correlation between the Shanghai Composite Index and interest rates, exchange rates, and then established a regression model between stock index returns and interest rates, exchange rates, and through the cointegration test shows that there is a long-run equilibrium relationship between interest rates and exchange rates and stock prices.

From the above literature summary, it can be seen that in the previous research, whether it is the impact of exchange rate or interest rate on stock prices, the correlation is time-varying and an uncertain correlation, as shown by different relationships within different research sample periods. Based on this, the author conducted an empirical study with the latest data from 2010-2024 to verify the specific direction of the impact of exchange rate and interest rate on stock prices during this sample period.

2.2 Analysis of the current situation of the stock, currency and bond markets

With the continuous development of China's capital market, its theoretical research is becoming more and more mature. Based on the fact that the writing of this paper mainly starts from the exchange market, bond market and stock market, so the author explores the current situation of the three markets and their interrelationships respectively, and further improves on the existing theoretical basis. Based on the timeliness of the research and the innovation of data, it is necessary to deepen the understanding of the current situation of the research factors first.

### 2.2.1 Analysis of the RMB exchange rate market

China's exchange rate system has been marketized from the fixed exchange rate in the early days to the flexible floating exchange rate nowadays. From 2010 to 2014, the exchange rate of RMB against the US dollar showed a trend of steady appreciation. During this period, China's rapid economic development, sustained growth in export trade and expanding trade surplus became important supporting factors for RMB appreciation. At the same time, China actively promoted the internationalization of the RMB by expanding the scope of its cross-border use and establishing a multilateral currency swap mechanism, which boosted the demand for the RMB in the international market. During this period, the RMB exchange rate against the US dollar gradually rose from 6.83 to around 6.15, an appreciation of about 10%. This appreciation trend not only reflected the strong performance of the Chinese economy and policy support, but also signaled an increase in the influence of the RMB in the global financial system.

When the RMB joined the basket of currencies as scheduled in 2015, the Central Bank further reformed the pricing rules for the RMB mid-price, and market makers quoted the exchange rate with reference to the RMB exchange rate price in the Hong Kong offshore market on the previous day in addition to the RMB exchange rate ratio of the RMB to the basket of currencies, which was more effective in avoiding violent shocks caused by speculative sentiment in the market, and provided a favorable capital market development in China. monetary environment for the development of China's capital market.

Thereafter, from 2016 to 2019, the RMB exchange rate was characterized by two-way fluctuations during this period. Factors such as trade friction between China and the United States, the trend of the U.S. dollar, and domestic economic policies combined to cause the exchange rate to appreciate and depreciate at times.

From 2020 to 2021, the Covid-19 epidemic had a profound impact on the global economy, but China's exchange rate against the US dollar appreciated significantly during this period thanks to rapid epidemic control and economic recovery. This appreciation process was driven by multiple factors. First, China became the first of the world's major economies to recover, with an economic growth rate of 2.3% in 2020, the only major economy to achieve positive growth. This has made the global capital markets more bullish on the outlook of the Chinese economy, and a large amount of international capital has flowed into the Chinese stock and bond markets, increasing the demand for the RMB. In addition, China's strong export performance, particularly in the areas of medical equipment, epidemic prevention supplies and electronics, and widening trade surpluses further supported RMB appreciation. Second, the U.S. dollar index weakened significantly in 2020. The Federal Reserve's ultra-loose monetary policy in response to the epidemic, including interest rate cuts and a large-scale asset purchase program, led to abundant dollar liquidity and depressed the dollar exchange rate. This created a favorable external environment for RMB appreciation.

From 2022 to 2024, the U.S.-China exchange rate experiences significant volatility, mainly driven by U.S.-China interest rate differentials, monetary policy divergence, and differences in economic fundamentals.

In 2022, the Fed opened aggressive interest rate hikes due to inflationary pressures, with a cumulative 425 basis points of interest rate hikes throughout the year, pushing the U.S. dollar index to strengthen significantly. Meanwhile, China's economic growth slowed down, repeated epidemics in many places, export growth declined, and the RMB depreciated against the US dollar to more than 7.2, with the market forming a long-term depreciation expectation for the RMB. 2023 saw a continuation of the depreciation pressure on the RMB, with the exchange rate approaching the 7.3 mark, mainly due to widening interest rate differentials between the US and China and weakening domestic exports. However, in the second half of the year, the PBOC curbed excessive volatility through interventions (e.g., foreign exchange market regulation and raising the foreign exchange reserve requirement ratio), and the exchange rate stabilized. In 2024, despite the Federal Reserve's pause in rate hikes but maintaining high interest rates, the dollar remained strong. China, on the other hand, stimulated economic growth through loose monetary policy, and RMB depreciation pressure gradually eased. As the domestic economy recovered and policies took effect, the RMB performed more stably in the second half of the year, with the exchange rate remaining in the 7.2-7.3 range.

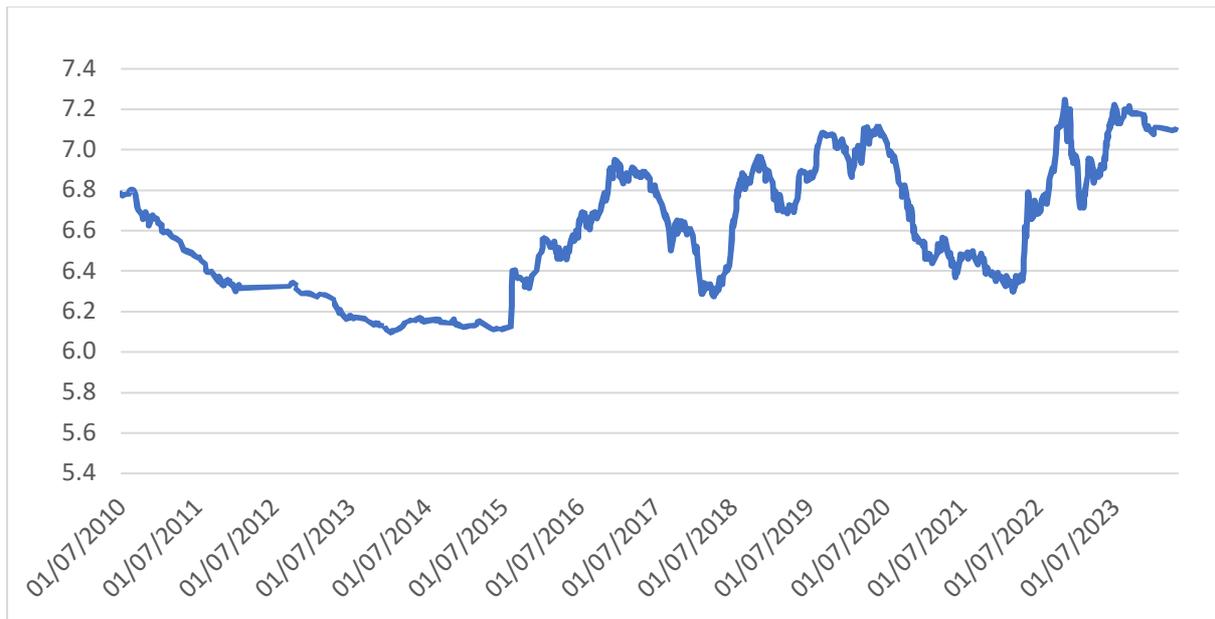

*Figure 1  2010-2024 USD-CNY Exchange Rate*

2.2.2    Analysis of domestic and international bond markets

The bond market, as an important sub-market acting as a fixed income player in the capital market, has been highly favored by international lobbyists, especially at a time when market risk aversion is on the rise. As of the end of August 2024, the custodian balance of China's bond market stood at RMB 167.9 trillion, according to data. Globally, China's bond market is the second largest, after the United States. The U.S. bond market is over $51 trillion, accounting for 39% of the global bond market.

From 2010 to 2014, China's bond market developed rapidly, with the balance under custody growing from RMB 20 trillion to nearly RMB 40 trillion, the market structure becoming more and more perfect, and the scale of issuance of government and corporate bonds expanding steadily, especially the issuance of municipal bonds and local government bonds becoming a special feature. In addition, the advancement of RMB internationalization has driven the rise of the offshore RMB bond market ("Dim Sum Bonds"). In the international bond market, the Fed's quantitative easing policy dominated the low-interest rate environment, global bond issuance rose sharply, especially in emerging market countries with the help of low-cost financing, and the international green bond market began to sprout at this stage, reflecting the initial concern for sustainable development.

Since 2014, the United States has been shouting interest rate hikes in the international market, which may imply some kind of political purpose, but it is undeniable that, regardless of whether or not the interest rate hike is ultimately successful, it has already stirred up waves in the global capital market. at the beginning of 2015, the yield on the U.S. 10-year treasury bond once fell below the 2% barrier, and then ushered in the counter-trend rebound pattern at the beginning of February, and then repeated the pattern of up and down oscillations, and in June 2015-July 2015, the yield on the U.S. 10-year Treasury bond once fell below 2%, and then rebounded against the trend, which was repeated continuously. During the period of June 2015-July 2015, it reached a great value for the whole year, touching a peak of 2.5%. the election of Donald Trump in the United States in October 2016, and his risk-loving style of action drove the rise of uncertainty in the global market, which in turn contributed to the warming of risk aversion, and stimulated the strengthening of the U.S. bond yield.

Comparatively speaking, Japan and Europe have yet to fully emerge from the shadow of the previous financial crisis, and have resolutely implemented negative interest rate policies against the backdrop of a sluggish economy. Although unlike the earlier QE policy as a helicopter to spread money to the surging, but the implementation of negative interest rate policy is also enough to ripple from time to time. In the negative interest rate policy environment, the market is expected to guide interest rates all the way down. At the same time, Brazil, India and other emerging economies also followed in the footsteps of China's economic downturn risk, the fundamentals are difficult to support, long-term bond rates downward channel has emerged.

In 2020, after the outbreak of the Corona epidemic, the U.S. government launched a large-scale fiscal stimulus plan, the Federal Reserve to adopt ultra-loose monetary policy, including significant interest rate cuts and asset purchases, resulting in a surge in the issuance of treasury bonds. 2020 to 2022, the U.S. Treasury market size from about 23 trillion U.S. dollars to expand to 27 trillion U.S. dollars. 2022, the Federal Reserve to cope with the high inflation to open the rapid interest rate hiking cycle, interest rates from near-zero level rapidly increased to 5 percent. rapidly increase from near-zero levels to above 5%. This move drove up Treasury yields while causing bond prices to fall, triggering liquidity pressures in the market and potential debt service risks.2023 saw a spate of regional bank failures in the U.S., exacerbating concerns about the stability of the Treasury market, with demand for short-term Treasuries surging in particular, and long-dated Treasuries coming under selling pressure. In 2023, the Treasury market expanded to $27 trillion as a safe-haven from inflation.

Nonetheless, the Treasury market, as a safe-haven asset, has attracted a large number of domestic and foreign investors. In early 2024, U.S. household holdings of Treasuries reached an all-time high of more than $2.4 trillion, reflecting investor confidence in Treasuries as a

stabilizing investment tool. However, over the long term, the resilience of the Treasury market and the trend in interest rates still face uncertainty as U.S. government debt continues to expand.

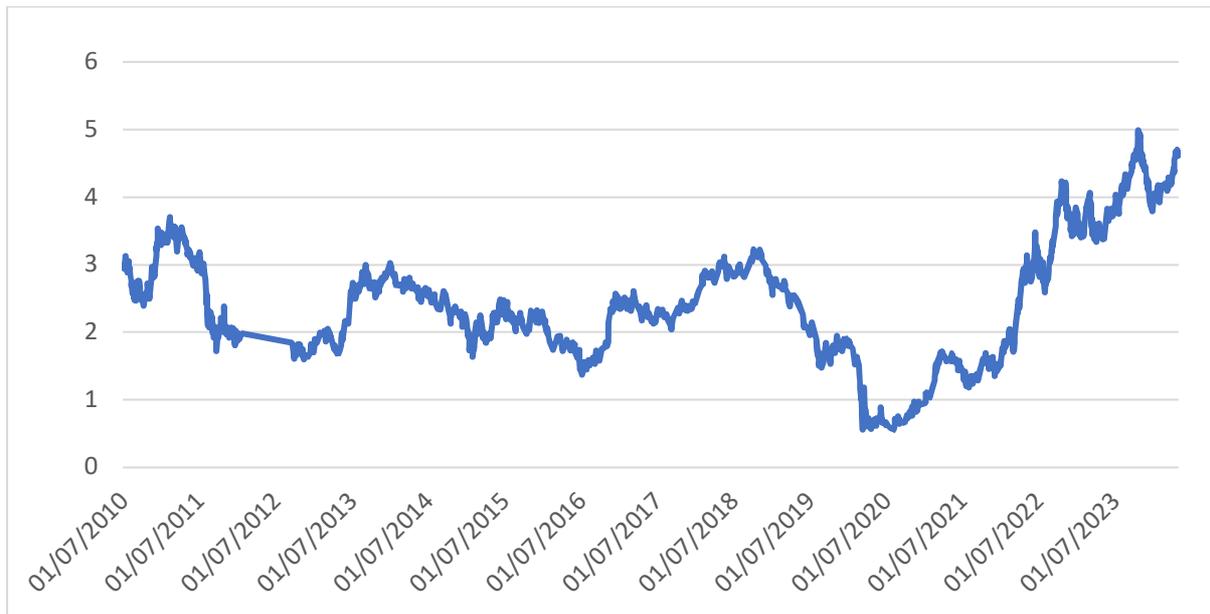

Figure 2  2010-2024 U.S 10-year Treasury Bond Yield

Compared with the U.S. bond market, 2015 to 2019 was a period of rapid growth in the size of China's treasury bond market. By the end of 2019, the size of China's treasury bond market reached approximately RMB 86 trillion, almost doubling from approximately RMB 47 trillion in 2015, with an average annual growth rate of over 15%. The inclusion of Chinese treasury bonds in major international bond indices, such as the Bloomberg Barclays Global Aggregate Index (2019) and the JPMorgan Government Bond Index, has boosted the entry of foreign investors. Between 2015 and 2019, the balance of foreign investors' holdings of Chinese bonds grew from about RMB 800 billion to nearly RMB 2 trillion, and the proportion of foreign investors increased from 1.7% to 2.5%, making treasury bonds the main investment underlying for foreign investors. In addition, China's central bank has maintained ample market liquidity through open market operations and medium-term lending facility (MLF) and other tools, risk management mechanisms have been gradually improved, and China's treasury bond yields are generally stable, with a high premium over developed country treasury bond yields, which has attracted more investors.

After the 2020 epidemic, China's treasury bond market experienced structural optimization and became an important financial instrument to cope with economic downward pressure. At the beginning of the epidemic, the PBOC implemented an accommodative monetary policy to stimulate the economy, and treasury yields briefly declined.2022 After 2022, as global inflation rose and monetary policies in major economies tightened, China's treasury bond market remained relatively stable, with yields rising slightly, demonstrating strong resilience. During the epidemic, the Chinese government increased fiscal spending to stabilize the economy, driving large-scale issuance of treasury bonds and local government bonds. By the end of 2023, the balance under custody in China's bond market exceeded RMB 140 trillion, of which treasury bonds accounted for a significantly higher share and became an important part of the market. Specialized local government bonds have played an important role in infrastructure development. As of 2024, the balance of Chinese bonds held by foreign investors reached RMB 3.6 trillion, with treasury bonds being the main allocated asset, accounting for more than 60% of foreign holdings. This trend reflects the international attractiveness of Chinese treasury bonds as a stabilizing asset.

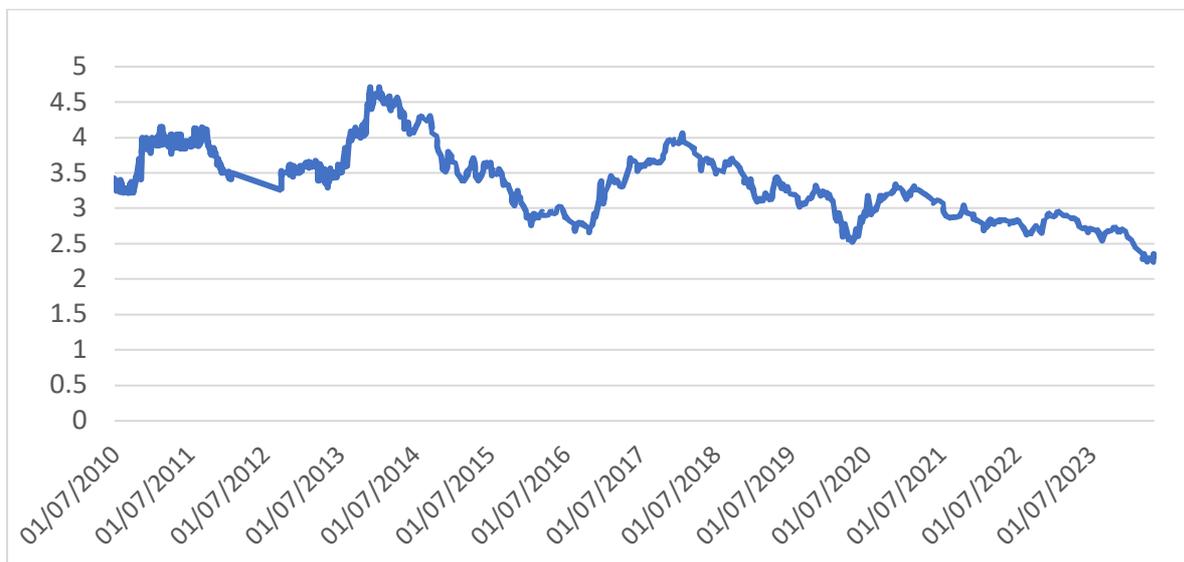

*Figure 3  2010-2024 Chinese Government 10-year Bond Yield*

2.2.3  Analysis of domestic and foreign stock markets

Whether it is a developed country represented by the United States or an emerging market country represented by China, the stock market can be seen as a barometer of a country's economic development. Behind the strong financial strength of the United States has a mature capital market as a strong support, institutional investors as the main stock market is one of the signs of maturity. Although China's A-share market is still dominated by retail investors, with the acceleration of a series of internationalization measures such as the Shanghai-Lunan Tong and the addition of the RMB to the SDR currency basket, as well as China's determination to never stop reforming, China's stock market will surely stride towards maturity in an unbiased manner, as if it were walking on a flat track.

The U.S. stock market, from 1929 to the present, the U.S. S&P index and EPS have grown by 5.2% and 5.1% respectively, and the PE has been maintained at around 19 times. A study shows that 60% of the bull market in the US stock market is due to profitability. Evidently, good profitability of listed companies is also a good medicine for the development of mature capital market.

After the financial crisis in 2008, the U.S. economy gradually recovered, and the stock market also rebounded. The S&P 500 Index rose from approximately 1,100 points in 2010 to approximately 2,050 points by the end of 2014, representing an average annual growth of approximately 16%. The growth during this period was primarily due to improved corporate earnings and quantitative easing by the Federal Reserve.

The stock market generally trended upward during the 2015-2019 period, but experienced volatility. in 2015 and 2016, the market experienced a brief correction due to the global economic slowdown and the decline in oil prices. However, in 2017 and 2018, the stock market rose strongly, buoyed by corporate tax reform and economic growth. in 2019, the market continued to grow despite trade tensions.

In early 2020, the Corona epidemic led to a sharp decline in the stock market, with the S&P 500 hitting a low of about 2,237 points in late March. However, thanks to government fiscal stimulus and Fed easing, the market rebounded quickly. By the end of 2021, the S&P 500 surpassed 4,700, an all-time high.

Inflationary pressures rise in 2022 as the economy recovers. The Federal Reserve began to raise interest rates to combat inflation, leading to increased market volatility, but in the second half of 2023, the market gradually rebounded as inflation moderated and economic data improved. In 2024, technology stocks continued to lead the market, with companies in the artificial intelligence and new energy sectors in particular outperforming. However, traditional sectors lagged relatively and the market diverged. As of April, the S&P 500 index stood at approximately 5,254, a new closing high, with a cumulative gain of 10.16% since 2024.

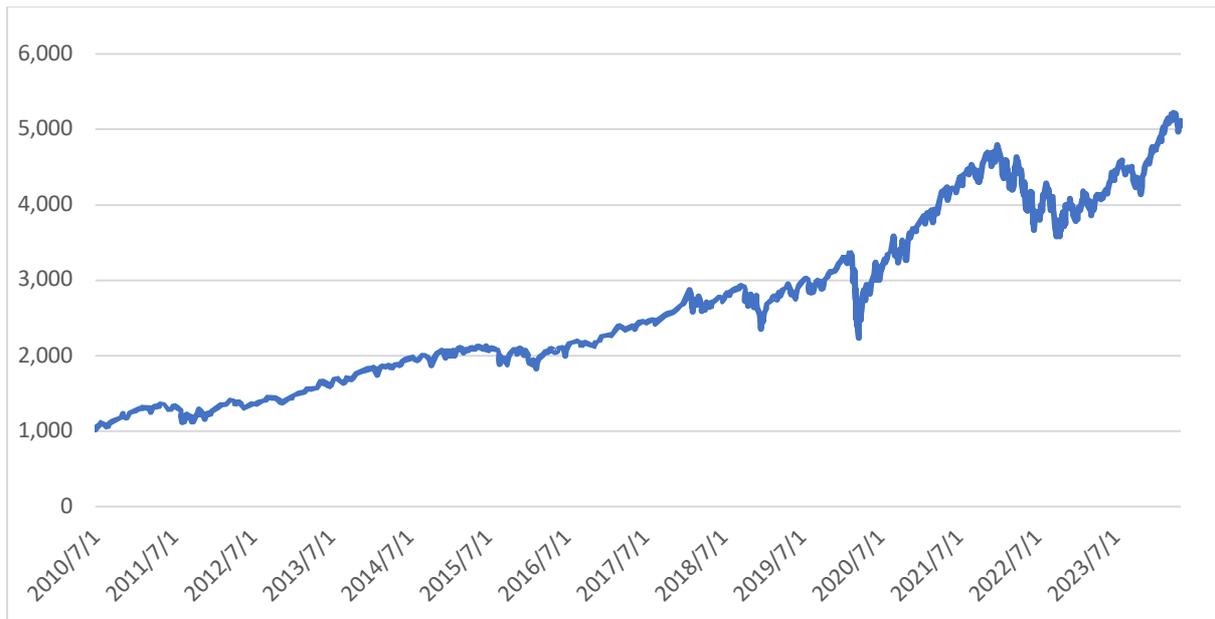

*Figure 4  2010-2024 S&P 500 Index*

In comparison, from 2000 to now, China's SSE Composite Index and EPS compound growth rate is 4.4% and 11.5% respectively, while PE has fallen from 43 times to 16 times. Some research shows that 90% of the transition between bull and bear markets in China's A-share market comes from valuation fluctuations rather than listed companies' profitability, which shows that the good development of China's capital market lacks the support of solid financial fundamentals, and the relevant decision makers and supervisors should pay attention to the monitoring of listed companies' profitability.

During the period from 2010 to 2014, the Shanghai Composite Index (SSE) showed a generally oscillating trend. at the beginning of 2010, the index was around 3,100 points, and then fluctuated between 2,000 and 3,000 points due to the slowdown of the domestic economic growth rate and the impact of policy regulation. in the second half of 2014, the index began to rebound as a result of favorable policies and improved market expectations. the stock market has been on a rapid rise since the beginning of 2010, and the stock market is still in the process of recovering.

In the first half of 2015, the stock market witnessed a rapid rise, with the SSE index exceeding 5,100 points in mid-June. However, a sharp market correction ensued, with the index falling sharply within a short period of time, back down to approximately 3,500 points by the end of 2015. This round of volatility raised market concerns about regulation and risk control. The central bank released liquidity and bled the stock market through multiple money market tools such as quota and interest rate cuts, open market operations, medium-term lending facilities and reverse repos. By January 2016, the performance of A-shares did not see a substantial rebound, but continued the decline of the second half of 2015, falling all the way to 2,655.66 points on January 28. after 2016, the stock market consistently fluctuated above and below 3,000 points, and market sentiment was relatively cautious. in 2018, as the trade friction between the United States and China intensified, the market was under pressure, with the index falling below 2,500 points at one point. in 2019, the market was under pressure, with the index falling below 2,500 points at one point. In 2019, the market gradually recovered as trade negotiations progressed and policies supported.

In early 2020, the stock market fell briefly due to the outbreak of the Xin Guan epidemic, but the index rebounded quickly, led by policy stimulus and economic recovery. in 2021, the market continued its upward trend, with outstanding performance in sectors such as science and technology and pharmaceuticals. in 2022, due to the impact of the global economic slowdown and the recurring domestic epidemics, the market adjusted, with the index fluctuating around the 3,000-point mark.

In the early part of 2024, China's stock market faced greater downward pressure, with the Shanghai and Shenzhen indices repeatedly hitting new stage lows, and market confidence was low. In order to boost market confidence, the regulators introduced a series of policy measures, including lowering of interest rates, supporting the development of the real estate market, and increasing support for scientific and technological innovation. As of April 2024, the SSE index oscillated around 2,500 points, and the overall market valuation was at a historically low level.

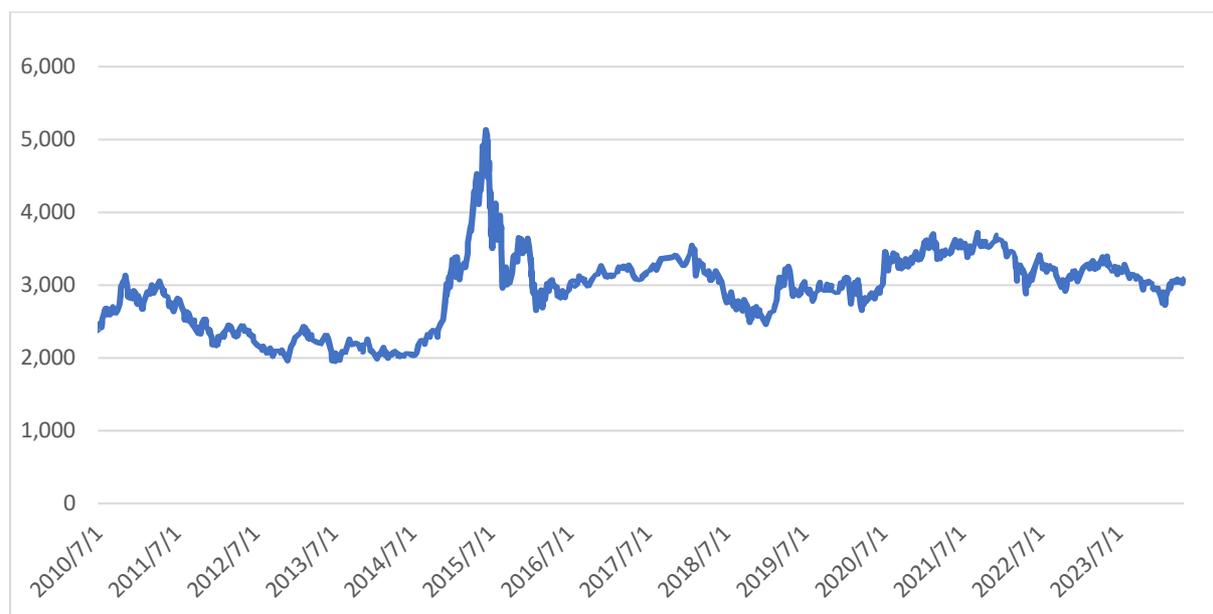

*Figure 5  2010-2024 Shanghai Composite Index*

# 3 Analysis of relevant theoretical foundations

## 3.1 Theoretical basis for the impact of bond yields on the stock market

### 3.1.1 Flow-oriented theory

In the era of economic integration and trade globalization, stock market and exchange market, as two important branches of capital market, have increasingly significant correlation. Western scholars Durnbush and Fisher put forward a more mature flow-oriented theory to study the relationship between the two in 1980.

The theory measures from the macro dimension and believes that there exists a positive causal relationship from the exchange rate to the stock price, emphasizing the current account or trade balance. This is manifested in the depreciation of a country's currency when the level of its exchange rate under the direct markup method increases. Whereas the exchange rate is the main factor affecting international trade, a rise in the exchange rate drives exports and dampens imports, which in turn leads to a trade surplus and GDP growth. With strong economic fundamentals as a support, there will be a positive spillover effect on corporate cash flow and stock price, so the exchange rate rise will push up the stock price, the two have a positive correlation. The transmission mechanism of the exchange rate on stock prices can be simplified as: Exchange rate rise → exports increase, imports decrease → trade surplus → GDP increase → stock prices rise

However, some scholars have questioned the flow-oriented theory, arguing that exchange rate changes affect the real value of the asset portfolios of domestic companies and multinational corporations, and thus affect the rise and fall of stock prices. For example, assuming that the RMB exchange rate has a rising momentum, the amount of foreign exchange collection of China's export-oriented enterprises will decline, which in turn will lead to a decrease in corporate profits and a decline in the company's share price.

It can be seen that the channels and results of the exchange rate impact on stock prices have not formed a unified opinion. Practical experience also shows that the impact of exchange rate risk on stock price varies with different underlying assets. Based on the historical performance of China's A-share market and the RMB exchange rate, resource stocks and industrial stocks react very differently to exchange rate fluctuations. During periods of RMB appreciation, the prices of industrial stocks tend to move upward, while during periods of RMB depreciation, the prices of resource stocks are firmer.

### 3.1.2 IS-LM-IA model

The IS–LM model, or Hicks–Hansen model, developed by John Hicks in 1937, shows the relationship between interest rates and output in the short run in a closed economy. The AD-IA, or aggregate demand–inflation adjustment model, developed by David Romer in 2000, builds on the concepts of the IS–LM model, essentially in terms of changing interest rates in response to fluctuations in inflation rather than as changes in the money supply in response to changes in the price level.

When stock prices rise, people's wealth increases due to the wealth effect, which in turn leads to an increase in consumption. Similarly, when the stock price falls, the market value of the enterprise will fall, and under Tobin's Q theory, it is more profitable to make market purchases than replacement costs, so enterprise investment will also fall. The inflation-adjusted (IA) curve can also be viewed as a Phillips curve.

The investment-saving curve is defined by the equation:
$$Y = C(Y - T(Y)) + I(i, S) + G + NX(q, Y)$$

The liquidity-money curve is defined by the equation:
$$M/P = L(Y, S, r)$$

The inflation-adjustment curve is defined as:
$$\pi = \pi^* + \lambda * (Y - \bar{Y})$$

where $Y$ denotes the level of national income, $C$ denotes consumption, $S$ denotes stock prices, $I$ denotes investment, $r$ denotes the level of interest rates, $G$ denotes government purchases, $NX$ denotes net exports, $q$ denotes the level of the real exchange rate, $M$ denotes the nominal supply of money, $P$ denotes the level of prices, $L$ denotes the demand for money, $\pi$ denotes inflation, $\pi^*$ denotes core inflation and $\lambda$ is a positive parameter that reflects how rapidly inflation responds to departures of output from its natural rate, and $\bar{Y}$ denotes the equilibrium level of national income.

Taking the real exchange rate $q$, the income $Y$ and the inflation rate $\pi$ as endogenous variables, the derivative of the stock price with respect to the exchange rate change is obtained by solving the above three equations with full differentiation:

$$\frac{\partial S}{\partial q} = -\frac{Ly * \phi q}{Ls * (1 - \phi y) + Ly * \phi s}$$

where $\phi s$, $\phi y$, and $\phi q$ represent the partial derivatives of the aggregate demand function with respect to the stock price, income, and the level of the real exchange rate, respectively, and $Ls$ and $Ly$ respectively represent the partial derivatives of the money demand function with respect to stock price and income. According to the theoretical knowledge of economics, it is known that all other things being equal, when the stock price rises, income increases, or the real exchange rate rises, respectively, the aggregate demand of the society increases, so $\phi s$, $\phi y$, and $\phi q$ are all positive. Similarly, the transactional demand for money is also positively related to the level of income, so $Ly$ is also positive. It follows that the interrelationship between the exchange rate and stock prices depends mainly on the sign of $Ls$, i.e., the relationship between stock prices and the demand for money. And stock prices affect money demand through various channels such as wealth effect, risk diversification effect, trading effect and substitution effect, with different effects affecting in different directions.

On the one hand, under the substitution effect, if the stock price rises, then investors prefer to hold stocks rather than money, so the less money is demanded, and the two move inversely, with $Ls$ being negative. On the other hand, under the wealth effect, transaction effect and risk diversification effect, if the stock price rises, the investor's wealth increases and the currency needed to carry out stock buying and selling transactions increases, so the more the demand for currency, and the two are positively related, with a positive value for $Ls$.

According to the formula for the derivative of the exchange rate to the stock price, it can be seen that there is a negative correlation between the exchange rate and the stock price when $Ls$ is positive. Whereas, when $Ls$ is negative, there is a positive correlation between the exchange rate and stock prices if the substitution effect of stocks on currencies is large enough.

It can be seen that the sign of $Ls$ can be positive or negative, which leads to an ambiguous relationship between the exchange rate and the stock price, which may be positively or negatively correlated.

   3.2 Transmission mechanism of exchange rates to stock prices

Interest rates are an important tool in the implementation of monetary policy by a country's central bank. Typically, when a central bank implements a tight monetary policy that reduces the money supply and thus increases market interest rates, it indicates that the market at that time is experiencing an overheated economy, a booming stock market and inflated asset prices. On the contrary, when the central bank adopts an expansionary monetary policy, which supplements market liquidity and thus leads to an increase in interest rates, it indicates that the market at that time is experiencing an economic depression and stock prices are going down. Therefore, in China's capital market, interest rates and stock prices are significantly correlated as representative indicators of the bond and stock markets, respectively. However, market interest rates under different systems do not have the same effect on stock prices. Under the floating interest rate system, the interest rate is only the intermediary target of the central bank for monetary policy regulation and control, while under the fixed interest rate system, the interest rate is the direct target of the central bank for monetary policy regulation and control. Therefore, under the fixed interest rate system, the impact of interest rates on stock prices is more significant. In addition, the theoretical community also exists a variety of interest rate on stock price transmission mechanism, different transmission mechanisms, interest rates through different channels directly or indirectly affect the stock market price, the author mainly summarizes the total social supply and demand effect, the social wealth effect and asset substitution effect of three kinds of transmission mechanisms.

### 3.2.1 Social Aggregate Supply and Demand Effect

Under this transmission mechanism, when interest rates rise, on the one hand, the cost of borrowing increases, so the opportunity cost of investing through borrowing rises, the expected rate of return falls, and investment decreases. On the other hand, it is more profitable for residents to save relative to consume because saving yields more returns on deposits, so the opportunity cost of consuming also rises and consumption decreases. In contrast, in a four-sector open economy, aggregate social demand consists of investment, consumption, government purchases and net exports. When the rise in interest rates leads to a decrease in both investment and consumption, aggregate social demand decreases, the market environment deteriorates, market supply is affected by pessimistic market expectations and demand decreases in tandem, the production and operation environment of enterprises deteriorates, and their business performance is hampered, ultimately leading to a decline in stock prices. On the contrary, when the interest rate falls, the opportunity cost of investment and consumption are reduced, the enterprise investment and consumption demand increases, the market environment is good, the social aggregate demand and aggregate supply rises at the same time, the enterprise production and operation environment improves, the operating performance improves, and ultimately lead to the rise in stock prices.

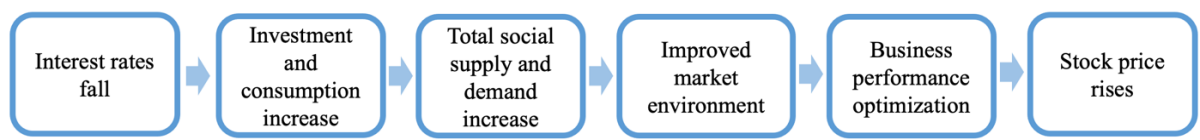

*Figure 6  The transmission mechanism of interest rate to stock prices under Aggregate Supply and Demand Effect*

However, the negative correlation between interest rates and stock prices does not always hold under the aggregate social supply and demand effect. If more relevant factors are taken into account, a rise in interest rates may also lead to a synchronized rise in stock prices. This is because market participants such as firms and investors usually focus on the size of the nominal interest rate and ignore the impact of the price level on the real interest rate. Therefore, when the level of nominal interest rates rises, but the price level of the more violent rise, the real interest rate is not rise but fall, so the opportunity cost of investment and consumption also fell, and then the total demand and supply of society increased, the business performance of enterprises improved, led to the rise in stock prices. In addition, enterprises in the investment and consumption decisions, not only by the level of interest rates only, but a combination of other factors also measured, may be the rise in interest rates cannot lead to a fall in investment and consumption. Therefore, to summarize, the correlation between interest rates and stock prices is not clear and may be either positive or negative.

### 3.2.2 Social Wealth Effect

Under the social wealth effect, the transmission of interest rates to stock market prices is mainly indirect through the intermediate element of total social wealth, usually preconditioned by a high savings rate. According to the principle of supply and demand in economics, the price change of any commodity or substance is due to the change of supply and demand. China is the world's largest saving country, especially China's rural residents, the excess cash in hand will firstly consider to deposit it in the bank in order to collect interest and gain income, lack of moderate investment concept. Therefore, when the interest rate rises, residents deposited in the bank savings will get more deposit interest income, which in turn leads to an increase in total social wealth. If the amount of stock investment in society as a whole as a proportion of total wealth remains unchanged, then when the overall wealth of society increases, the demand for stocks subsequently increases. Under the condition of constant supply and increased demand, there is an increase in the price of shares. On the contrary, if the interest rate falls, then the interest income from bank deposits falls, which will inevitably lead to a decline in the total wealth of the society of a country with a high savings rate, and under the condition that the proportion of investment in stocks remains unchanged, the supply of stocks in the market remains unchanged, the demand decreases, and the price of the stocks subsequently falls. Therefore, theoretically, under this effect, interest rates and stock prices are positively correlated.

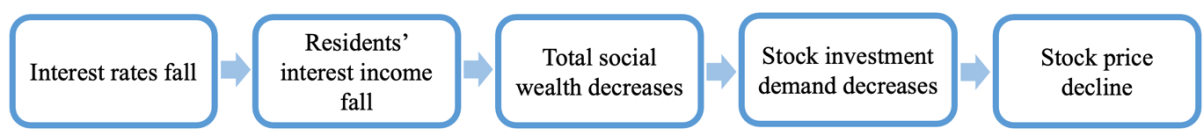

*Figure 7 The transmission mechanism of interest rate to stock prices under the Social Wealth Effect*

### 3.2.3 Portfolio Substitution Effect

American economist Harry Markowitz (1952) for the first time put forward the asset portfolio theory, used to explain the expected risk and expected return matching relationship. The theory comprehensively analyzes equity investments such as high-yield, high-risk stocks and fixed-income investments such as low-yield, low-risk bonds. Typically, risk-seekers will choose to invest more in equities in order to obtain a higher risk premium, while risk-averse investors will prefer fixed-income investments, such as bonds, in order to minimize possible losses. In the optimal asset portfolio, both risky assets such as stocks and risk-free assets such as state bonds need to be included, and qualified investors should allocate their funds rationally in the stock market and the bond market to maximize the value added.

The asset portfolio substitution effect is based on the asset portfolio theory mentioned above, which also confirms the "seesaw" relationship between the bond market and the stock market in China's capital market. Changes in interest rates indirectly affect stock market prices through capital investment in the stock and bond markets. Rational investors will reasonably match the investment ratio between the stock market and the bond market to maximize returns. First, when interest rates rise, bond prices fall and bond yields rise. At this time, the relative return on investment in stocks falls, so investors seeking to maximize returns will increase demand for bonds and decrease demand for stocks, resulting in a fall in stock prices. Conversely, when interest rates fall, bond prices rise and expected yields fall. At this point, investors will reduce the amount of money invested in bonds and increase the amount invested in stocks, which leads to an upward movement in stock prices. So there is a negative correlation between interest rates and stock prices under the substitution effect of asset portfolios.

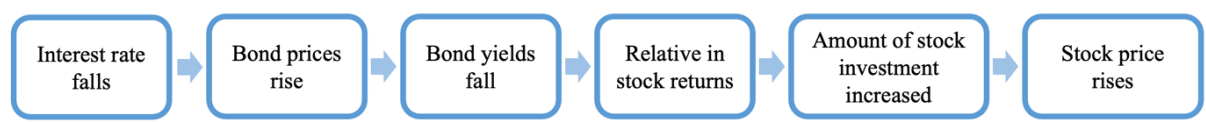

*Figure 8  The transmission mechanism of interest rate to stock prices under the Portfolio Substitution Effect*

However, the asset mix approach to the transmission of interest rates to stock prices is also flawed, which makes it less accurate. First, the effect only takes into account the substitutability between the bond market and the stock market, and ignores the substitutability between the stock market and other markets such as the money market and the foreign exchange market. China's capital market has multi-level attributes, so a fall in bond interest rates does not necessarily lead to a diversion of funds from the bond market to the stock market. Secondly, the portfolio substitution effect only aims at maximizing expected returns and ignores the goal of risk minimization. In reality, although the relative return of the stock market is very high, it is beyond the tolerance range of many risk-averse investors, who would rather buy risk-free or low-risk assets with lower yields than to take greater risks to obtain greater returns. Therefore, when bond interest rates fall, risk-averse investors may not reduce their investment in bonds and increase their investment in stocks, but choose to continue to buy low-risk bonds and give up high-yield and high-risk stocks. At this point, the negative correlation between the exchange rate and stock prices is no longer valid.

In addition, Keynes's liquidity preference theory also suggests that the effect of interest rates on stock price changes is not unique, and that it is necessary to compare the magnitude of interest rate adjustments with the magnitude of expected changes. Only when interest rates adjust faster than the expected change, the increase in interest rates will have the effect of raising stock prices. Similarly, when interest rate adjustment is slower than the expected change, the interest rate rise instead of the stock price to play a dampening effect. In the academic world, there are also policy signal effect, cost-benefit effect, stock intrinsic value effect and other theories to study the transmission mechanism of interest rates on stock prices, in the three transmission channels, interest rates and stock prices are negatively correlated with each other. To summarize, there is not a unique correlation between interest rates and stock prices, but rather different relationships under different transmission mechanisms.

### 3.3 Theoretical basis for the impact of exchange rate on the stock market

#### 3.3.1 Discounted Cash Flow Theory

The discounted cash flow theory, first proposed by Williams in 1938, is one of the most basic theories of market value. Under this theory, the time value of money is taken into account, and the principle of present value is utilized, which is a method of discounting future cash flows to the current period through a certain discount rate. It is not only suitable for analyzing the market value of stocks, but also for evaluating the value of enterprises, measuring the cost of financing, and making investment decisions.

For the stock held by the investor, its subsequent cash flow mainly comes from two aspects: the dividend income in the future holding period and the selling price when the stock is sold. If the investor chooses to sell the stock in the next n years, the current market value of the stock according to the discounted cash flow theory is:

$$PV = \frac{D_1}{(1+r)^1} + \frac{D_2}{(1+r)^2} + \frac{D_3}{(1+r)^3} + \cdots + \frac{D_n}{(1+r)^n} + \frac{TV}{(1+r)^n}$$

where $D_n$ denotes the future dividend or bonus on common stock in year $i$, $TV$ denotes the selling price of the stock after year $n$, $PV$ denotes the current market price of the stock, and $r$ denotes the discount rate, also known as the capitalization rate, which is usually based on the current market interest rate, i.e., the sum of the risk-free rate of return on the bond plus the market risk premium.

If the investor chooses to hold the stock in perpetuity, the dividend is the only source of future cash flow. This is the famous discounted dividend model. It can be seen that, for the stock market, the discounted dividend model is only a special form of the discounted cash flow model, and its special features are mainly manifested in two aspects: first, the future selling price $TV$ is equal to 0, and second, the future period is no longer $n$ periods, but infinity. Thus, the specific expression of the discounted dividend model is:

$$PV = \frac{D_1}{(1+r)^1} + \frac{D_2}{(1+r)^2} + \frac{D_3}{(1+r)^3} + \cdots + \frac{D_n}{(1+r)^n} + \cdots = \sum_{i=1}^{\infty} \frac{D_i}{(1+r)^i}$$

Under the discounted cash flow theory, both interest rates and stock prices are negatively correlated, i.e., when interest rates rise, the rate of compensation demanded by investors increases, the discount rate rises, which in turn leads to a decrease in the sum of the present value of future cash flows from the stock and a downward movement of the stock price. Similarly. When interest rates fall, stock prices rise in the opposite direction.

### 3.3.2 Capital Asset Pricing Theory

The Capital Asset Pricing Model (CAPM) was developed based on portfolio theory and capital market theory, and was proposed by American scholars William Sharpe, John Lintner, Jack Treynor and Jan Mossin. The main study is the relationship between the stock market yield and the yield of wind-uninsured treasury bonds.

Under the CAPM theory, the main factor affecting the requisite rate of return on a risky asset is the stock's $\beta$. $\beta$ measures the correlation between the return on that asset and the return on the market portfolio, and is mainly affected by the standard deviation of the stock itself, $\delta_i$, the standard deviation of the return on the market as a whole, $\delta_m$, and the correlation coefficient of the stock with the stock market as a whole, $L_{im}$.

In particular, $\beta$ varies in the same direction as $\delta_i$ and $L_{im}$ and inversely as $\delta_m$. The specific form is:

$$\beta = \frac{L_{im} * \delta_m * \delta_i}{\delta_m^2} = L_{im} * \frac{\delta_i}{\delta_m}$$

$\beta$ measures the magnitude of systematic risk that cannot be offset, with larger values indicating greater systematic risk. When $\beta = 1$, the systematic risk of the stock is consistent with the risk of the market portfolio. When $\beta = 0$, the stock has no systematic risk. Similarly, when $\beta > 1$, the systematic risk of the stock is greater than the risk of the entire market portfolio. Generally speaking, a reasonable range of values for β is between 0 and 1.

For a risky asset such as an equity, its expected return $E(R_i)$ can be expressed, on the one hand, as the Treasury bond return Rf plus the risk premium of the equity. The risk premium component is the product of the stock's $\beta$-value and the difference between the expected market rate of return, $E(R_m)$, and the risk-free rate of return, $R_f$. On the other hand, in conjunction with the market price of the stock, and in accordance with the methodology used by the SSE Composite Index to calculate the yield, the stock yield can also be expressed as the logarithmic form of the quotient of the current period's stock price and the previous period's stock price, multiplied by 100. The logarithmic form filters out the trendiness of the data, making it more accurate and reliable compared to the yield expressed in terms of the rate of growth of the stock price. The specific form is shown below:

$$E(R_i) = R_f + \beta_i * (E(R_m) - R_f) = lnP_t - lnP_{t-1}$$

It can be seen that the stock return is not only positively proportional to the $\beta$-value of the stock itself and the market yield, but also affected by the interest rate of the bond market, which is also positively correlated, i.e., the higher the interest rate of treasury bonds in the bond market, the higher the corresponding stock return. The stock return is positively and negatively correlated with the current day's stock price, $P_t$ and the previous day's stock price, $P_{t-1}$ respectively. Therefore, under the capital asset pricing theory, there is a positive correlation between the interest rate and the current stock price, and a negative correlation with the stock price of the previous day.

### 3.4 Transmission mechanism of interest rates to stock prices

In the context of an open economy, the intrinsic linkage between the foreign exchange market and the capital market is very significant, even if there is a certain amount of foreign exchange control, the linkage between the two is also relatively close, which is directly reflected in the transmission effect of the exchange rate on the stock price. Generally speaking, the main existence of competitiveness effect, income effect, cheap import effect and inflation effect.

#### 3.4.1 The Competitiveness Effect

Under the direct markup method, when a country's exchange rate rose, that is, when the country's currency depreciation, the competitiveness of the country's goods to strengthen, and thus exports to foreign countries increased. At the same time, due to the national currency in foreign markets, the purchasing power of the decline, and domestic due to currency depreciation led to pessimistic investor expectations, so that domestic enterprises to reduce investment, thus weakening the demand for imports. Under the conditions of export enhancement and export reduction, the country's trade surplus increased, foreign exchange reserves rose, strengthening the foundation of the country's economy, the formation of fundamental support for the stock market and a major positive, which in turn led to the country's stock prices rose. Therefore, under the competitiveness effect, the exchange rate on the stock price of the transmission performance of the two have a positive relationship.

The specific transmission mechanism is shown in the figure below:

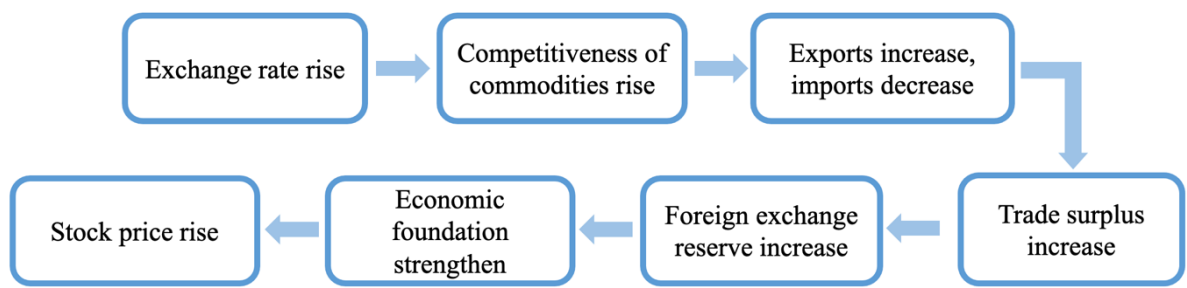

*Figure 9 The transmission mechanism of the exchange rate to the stock price under the Competitiveness Effect*

3.4.2 The Cheap Import Effect

Under this transmission mechanism, the rise in the level of a country's exchange rate indicates that the country's currency has depreciated, and its price level relative to the foreign market declined. As a result, the country's residents will be more inclined to spend at home, which will lead to an increase in the demand for local currency and a decrease in the demand for foreign currency. As a result, the central bank's foreign exchange reserves then increase, favoring the domestic economy and benefiting from rising stock prices. However, under the cheap imports effect, the depreciation of the domestic currency also indicates that the country's export prices are relatively lower, and the price of imports is relatively higher, which in turn leads to a decline in the country's nominal income, making the level of consumption lower, economic development is hindered, dragging the stock market downward. Therefore, under this effect, the transmission of the exchange rate to the stock price is mainly characterized by an uncertain correlation between the two.

The specific transmission mechanism is shown in the figure below:

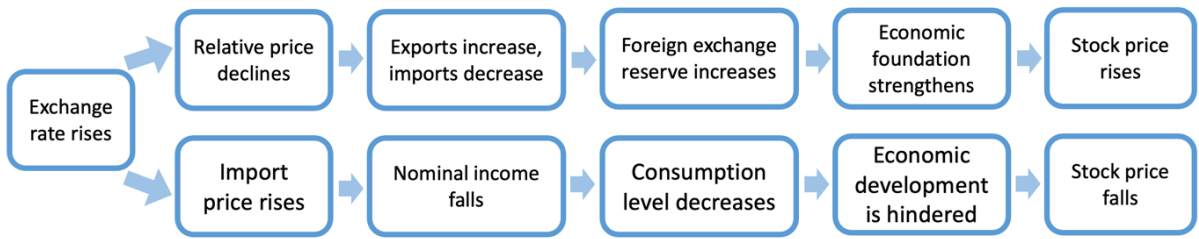

*Figure 10 The transmission mechanism of the exchange rate to the stock price under the Cheap Import Effect*

### 3.4.3 The Inflation Effect

When a country's exchange rate under the direct markup method rises, it means that the currency depreciates, which promotes the export of domestic commodities, so the supply of domestic commodities is reduced accordingly, prices rise in disguise, and inflation appears. At this point, under the Fisher effect, in order to keep the real rate of return unchanged, the nominal rate of return will rise with the inflation rate. As a result, the discount rate using the discounted cash flow model rises, which leads to a decline in stock valuation and a setback in stock prices. However, academics have also argued that using nominal interest rates directly as a criterion for the discount rate is not reliable. Rather, it is argued that when inflation rises, it leads to an increase in the price risk of future financial assets, and thus a rise in the rate of return on risk demanded by investors, which is what leads to a rise in the discount rate, a fall in company valuations and a fall in share prices.

In addition, another transmission mechanism of the inflation effect argues that when inflation occurs, due to rising prices, firms are concerned about future profitability and pessimistic market expectations emerge, which leads to a decline in company dividends, a fall in stock valuations, investors selling their holdings, and ultimately dragging stock prices downward. It can be seen that, under this effect, the exchange rate is ultimately indirectly transmitted to the stock price through the expectation of inflation, and there is a negative correlation between the two.

The specific transmission mechanism is shown in the figure below:

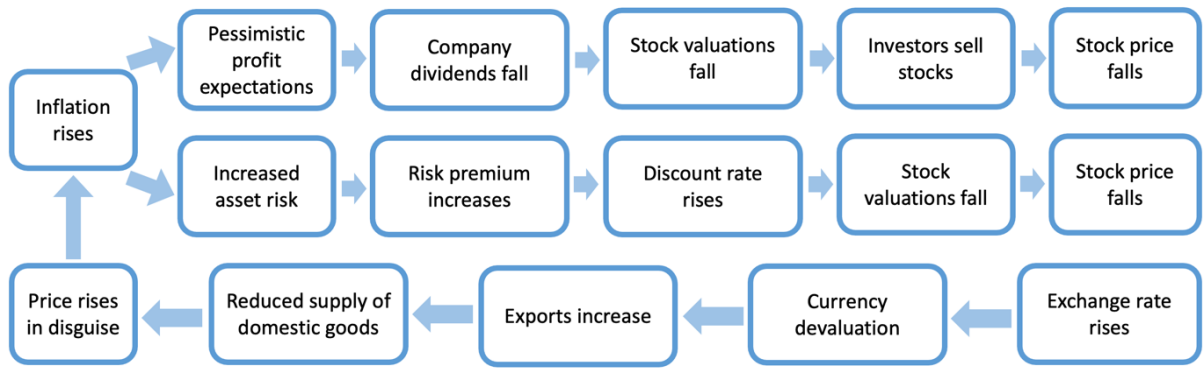

*Figure 11 The transmission mechanism of the exchange rate to the stock price under the Inflation Effect*

## 4   Methodology

### 4.1 Introduction to ARDL Model

Autoregressive Distributed Lag (ARDL) was proposed by Jorgenson in 1966, the core idea is that the model should not only consider the effect of the independent variable on the dependent variable at the same time, but also consider the effect of the lag of the dependent variable itself and the lag of the independent variable and add a lagged variable. As we all know, the lagged effect caused by the time difference factor is significant in the transmission between macroeconomic variables, so the ARDL model has more realistic significance in the study of economic problems. Moreover, the addition of lagged variables not only improves the model's goodness of fit, but also more accurately depicts the dynamic change and adjustment process of the whole economic phenomenon.

For the general form of autoregressive distributed lag model ARDL (n, m) has the following expression:

$$Y_t = C_0 + C_1 Y_{t-1} + C_2 Y_{t-2} + C_3 Y_{t-3} + \cdots + C_n Y_{t-n} + \beta_0 X_t + \beta_1 X_{t-1} + \cdots + \beta_m X_{t-m} + \varepsilon_t$$

$$= C_0 + \sum_{i=1}^{n}(C_i * Y_{t-i}) + \sum_{i=0}^{m}(\beta_i X_{t-i}) + \varepsilon_t$$

where $C$ ($C_0, C_1, C_2 \ldots C_n$) and $\beta$ ($\beta_0, \beta_1 \ldots \beta_m$) denote the correlation coefficients of the lagged variables with respect to the current dependent variable, i.e., the parameter vector.

$X_{t-i}$ denotes the lagged independent variable at lag *i*.

*n*, *m* denote the maximal lag order of endogenous variable *Y* and exogenous variable *X*, respectively, and $\varepsilon_t$ denotes the error term, which represents all other correlations that would affect the dependent variable but did not enter the model eventually and obeys the normal distribution $\varepsilon$ ($\varepsilon_1, \varepsilon_2 \ldots \varepsilon_{t-1}, \varepsilon_t$) ~ *N (0, $\sigma^2$)*, i.e.:

$$\text{E}(\varepsilon_t / Y_{t-1}, Y_{t-2} \ldots Y_{t-n}, X_t, X_{t-1} \ldots X_{t-m}) = 0, \ \text{S}(\varepsilon) = \sigma^2$$

When each correlation coefficient (, ) of the exogenous variables in the model is 0, the ARDL model is the Autoregressive Model (AR model), i.e.:

$$Y_t = C_0 + C_1 Y_{t-1} + C_2 Y_{t-2} + C_3 Y_{t-3} + \cdots + C_n Y_{t-n} + \varepsilon_t = C_0 + \sum_{i=1}^{n}(C_i * Y_{t-i})$$

When each correlation coefficient C (C0, C1, C2...Ci) of the endogenous lag variables Yt-n in the model is 0, the ARDL model is a Distributed-Lag Model (DL Model), i.e.:

$$Y_t = \alpha + \beta_0 X_t + \beta_1 X_{t-1} + \cdots + \beta_m X_{t-m} + \varepsilon_t = \alpha + \sum_{i=0}^{m}(\beta_i X_{t-i}) + \varepsilon_t$$

It can be seen that the AR model and DL model are only two special forms of ARDL model, while the ARDL model is a combination of the general forms of AR and DL model, so the ARDL model has a stronger generalization and a wider scope of application. In addition, the autoregressive distributed lag model has at least the following two advantages:

I. ARDL is a dynamic model. Compared with the static model, it considers the influence of time factor on endogenous variables, reflects the interrelationship between variables more directly and objectively, and is characterized by clarity, visibility, and simplicity. Generally speaking, the fit of the static model is low, and there is a large observation error in the sample observation value obtained through the survey. Moreover, when the model is used for forecasting, more forecasting needs are for the dynamic model.

II. The stationarity of the time series data is not required when using ARDL modeling. On the contrary, when using VAR, MR and ARIMA models to establish the functional relationship between variables, the time series variables need to ensure the stationarity of the series, eliminating the influence of time trends or random factors, but the ARDL model itself has already substituted the lagged time factors into the model during the modeling process, so the stationarity of the time series data is less demanding.

Through the preliminary elaboration of the ARDL model, it can be seen that the autoregressive distributed lag model can more accurately reflect the nature of economic operation in the study of economic problems with time lag effect, which is more representative and practical. It is not difficult to understand that it takes some time to digest the relevant economic policies or good news from the announcement to the transmission and the final effect. In particular, the implementation of China's monetary policy is generally considered to have a time window of at least 4-6 months before it has an impact on indicators such as GDP, which is a measure of economic development. For example, tightening monetary policy requires raising interest rates by reducing the money supply, then reducing investment by raising interest rates, and finally regulating the economy through the impact of investment on GDP to achieve its ultimate goal.

The more intermediate factors that need to pass through in the transmission process, the more significant the time lag effect will be, and the value of the ARDL model will be more prominent. In addition, when the variables do not satisfy the linear distribution lag model, a nonlinear distribution lag model will be used, which deepens the linear ARDL model by choosing the appropriate function form and reasonable treatment of variables for estimation.

### 4.2 Data Source and Variable Selection

Complying with the requirements of academic rigor and data accuracy, the thesis pays much attention to its authority in the selection of samples, and all the data come from the official website of the National Bureau of Statistics, the official website of the People's Bank of China, the official website of the Shanghai Stock Exchange, the Wind Economic and Financial Database, the Flush database and other authoritative data terminals, in order to more accurately describe the research topic and to reflect the essence of the economic phenomenon behind it.

In the selection of data, the author chooses the daily data of each variable. First of all, whether it is the daily data of the Shanghai Composite Index or the daily data of the RMB exchange rate, or the daily data of the national bond yields of various countries, all of them can be found directly on the official website of the People's Bank of China, so the data are more available. In addition, due to the fast-changing nature of the stock market, daily data is more valuable than monthly, quarterly and annual data, and better reflects the development and change patterns of the stock market. Moreover, the selection of daily data can obtain enough samples, which can better reveal the laws of economic operation, reduce the model error and improve the accuracy. Finally, based on the timeliness of the analysis, the author chooses the daily data of the relevant variables from 2010 onwards until the beginning of the writing date, i.e. April 2024.

For the selection of indicators of stock index performance, the Shanghai Composite Index of China's A-share market was chosen. First, the Shanghai Stock Exchange is the main stock market in China. Second, the SSE Composite Index is a comprehensive indicator that more accurately indicates the general market trend, weighting the vast majority of stocks in the market, and is the most authoritative indicator reflecting the general market trend.

In terms of the choice of variables for measuring the exchange rate, the exchange rate of RMB against the US dollar was selected. First of all, it is because the exchange rate of RMB to USD can be seen as a reference standard for other exchange rates, and it is also the exchange rate that investors in the market are most concerned about. Moreover, the sample is chosen to be the spot exchange rate rather than the mid-rate of the exchange rate in order to more fully consider market factors and fully reveal the most realistic level of the exchange rate under the effect of market supply and demand.

In terms of interest rate selection, the first choice is the 10-year long-term treasury bond yield, because the long-term treasury bond yield not only represents the level of nominal interest rate, but also the risk-free interest rate, the data is relatively stable, and it is also the basic index for setting other market interest rates, and the 10-year treasury bond yield of the United States is the most representative in the U.S. treasury bond market, which usually reflects the trend of the international capital flow. In terms of the choice of foreign treasury yields, the author uses the U.S. treasury yields as the main independent variable, combined with the exchange rate to comprehensively reflect the impact of international market factors on the performance of China's stock index. Finally, the samples are taken from the latest data from July 1, 2010 to April 30, 2024, a total of 3150 sample data, through a large sample in order to exclude the influence of randomness and establish accurate model relationships.

### 4.3 Variable construction and preliminary data processing

As can be seen through the above sample data selection, the article chose the U.S. 10-year Treasury bond yield as the main independent variable for modeling, because in the case of China's stock market running stationarily and performing well, on the one hand, investors have a relatively high degree of risk tolerance for China's stock market, and a large amount of money will be withdrawn from the U.S. Treasury market in order to seek for a higher rate of return. On the other hand, when U.S. Treasury yields fall, the market demand for U.S. dollars decreases, which in turn leads to a depreciation of the dollar and a relative appreciation of the yuan, in order to maintain the stability of the yuan exchange rate, the central bank will buy U.S. dollars in the foreign exchange market and sell the yuan, at which time the liquidity of the yuan in the market increases, so the inflow of funds into China's stock market increases, and the stock price further upward. Therefore, theoretically speaking, in the period of good performance of China's stock market, the U.S. 10-year Treasury market and China's A-share market mainly through the impact of capital inflows and outflows and the demand for U.S. dollars to show a negative correlation. It can be seen that the impact of U.S. Treasury bonds on the Shanghai Composite Index has a more mature theoretical system. Moreover, many scholars have already studied the relationship between U.S. Treasury yields and the Shanghai Composite Index. Ying Xiaoyun (2014) found that there is a two-way volatility spillover effect between the U.S. 10-year Treasury bond yield and the Shanghai Composite Index through the VAR-MGARCH-BEKK model, and found that the U.S. 10-year Treasury bond yield can have a significant impact on China's Shanghai Composite Index through the VAR model.

In order to fully demonstrate the impact of exchange rate, domestic and foreign bond yields on China's Shanghai Composite Index, the author conducted a study with the yield of the Shanghai Composite Index as the dependent variable, mainly based on the fact that the respective variables are all relative values, and cannot be directly modeled with the absolute value of the stock price as the dependent variable. On top of the existing research foundation, the author applies the principal component analysis method to the yield indicator. Among them, in order to suppress the trend characteristics of stock price, which is often a time-series data, the calculation of the rate of return is based on a logarithmic formula, rather than a simple growth rate indicator. Other control variables are not included in this paper because the model itself needs to consider the impact of lags of all variables, and the number of variables will be much higher after adding lags. Secondly, the sample data of the article is daily data, and many macroeconomic variables affecting the stock market, such as GDP, inflation rate, etc., do not have corresponding daily data, so the availability of data is not strong. Furthermore, the author reviewed many studies using the ARDL model, compared with the regression model that only considers the current period's impact, the ARDL model should not have too many independent variables is not conducive to the interpretation of the economic meaning of the main components. Finally, the specialized analysis of the main independent variables can also make the study more in-depth.

As a result of the above analysis, the dependent variable is the volatility of the return on the Shanghai Composite Index.

Indicators for each variable can be specifically expressed as follows: YIELD (Shanghai Stock Exchange daily yields): SSE daily returns; VOL (Yields Volatility): Daily volatility of the stock market; FXR (Foreign Exchange Rate): Daily USD-RMB exchange rate; CNB (China Bond Yield) Daily Chinese Government Bond yield; USB (United States Bond Yield): Daily US Treasury Bond yield. As shown in the table below:

*Table 1 Variable original design table*

| Variable Type | Variable Symbol | Variable Name | Source |
| --- | --- | --- | --- |
| --- | YIELD | SSE daily returns | |
| Dependent Variables | VOL | Daily volatility of the stock market | |

| | VOL(-n) | Daily volatility of the stock market (lagged variable) | Sourced from authorities such as the National Bureau of Statistics and the Wind database. |
|---|---|---|---|
| Independent Variables | FXR | Daily USD-CNY exchange rate | |
| | CNB | Daily Chinese Government Bond yield | |
| | USB | Daily US Treasury Bond yield | |

### 4.3.1 Chinese Stock Market Volatility Variables

In this paper, the Shanghai Composite Index (SSE) is chosen as an indicator of China's stock market. Firstly, the Shanghai Composite Index includes all the stocks listed on the Shanghai Stock Exchange, which is more comprehensive in terms of the number of stocks and industries than the constituent indexes such as Shenzhen 100 and CSI 300; secondly, the constituent indexes are selected from the top stocks in terms of liquidity and size in the Shanghai and Shenzhen markets, which represent the changes in the large- and mid-cap stocks, and the SSE Composite Index can reflect the market performance of large-, medium-, and small-cap stocks comprehensively. The SSE Composite Index can reflect the market performance of large, medium and small-cap stocks. Therefore, the SSE Composite Index is chosen as a representative of the Chinese stock market, which can reflect the changes of the Chinese stock market in a more comprehensive way.

#### 4.3.1.1 Construction of monthly stock market volatility

In this paper, stock market volatility is calculated according to the methodology of Zhang Zongxin and Wang Hailiang (2013) , and the monthly volatility is calculated as the volatility of the return of the SSE Composite Index during the month, which is given in the following formula:

$$mvol = \sqrt{\frac{\sum_{j=1}^{n}(r_j - \bar{r})^2}{n-1}}$$

Where $mvol$ denotes the stock market volatility; $\bar{r}$ is the calculated average stock market return; $r_j$ is the calculated daily log stock market return; and $n$ is the number of days in each month.

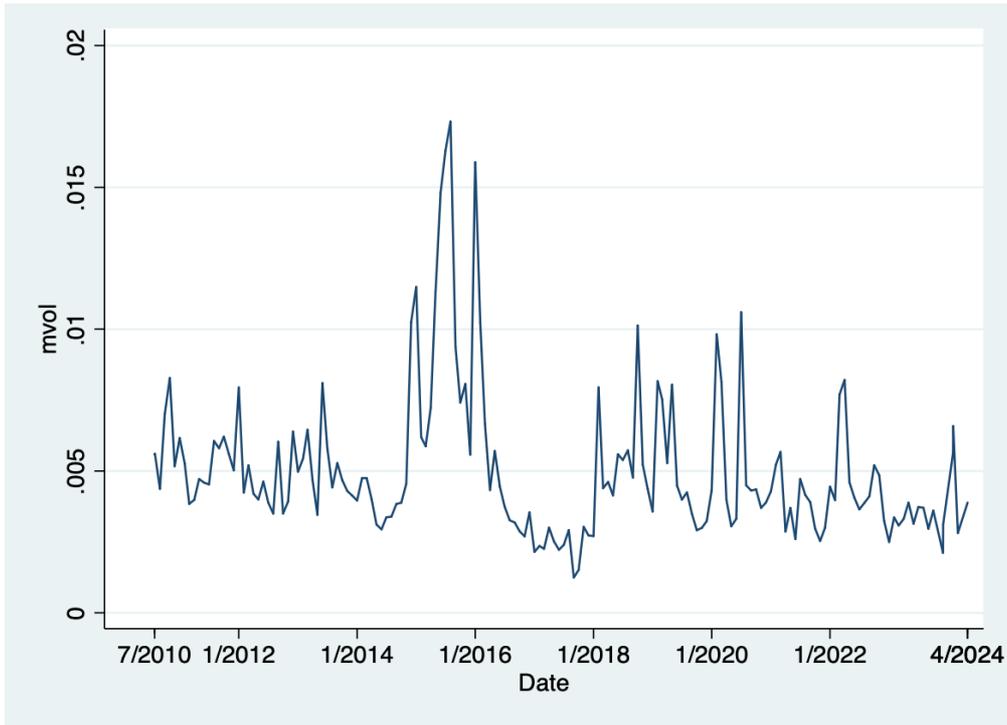

*Figure 12  Monthly Stock Market Volatility*

4.3.1.2 Construction of daily stock market volatility

In this paper, the GARCH(1,1) model is used to construct the daily volatility of the stock market, and the selected time period is from July 1, 2010, to April 30, 2024 to construct the daily volatility of the stock market.

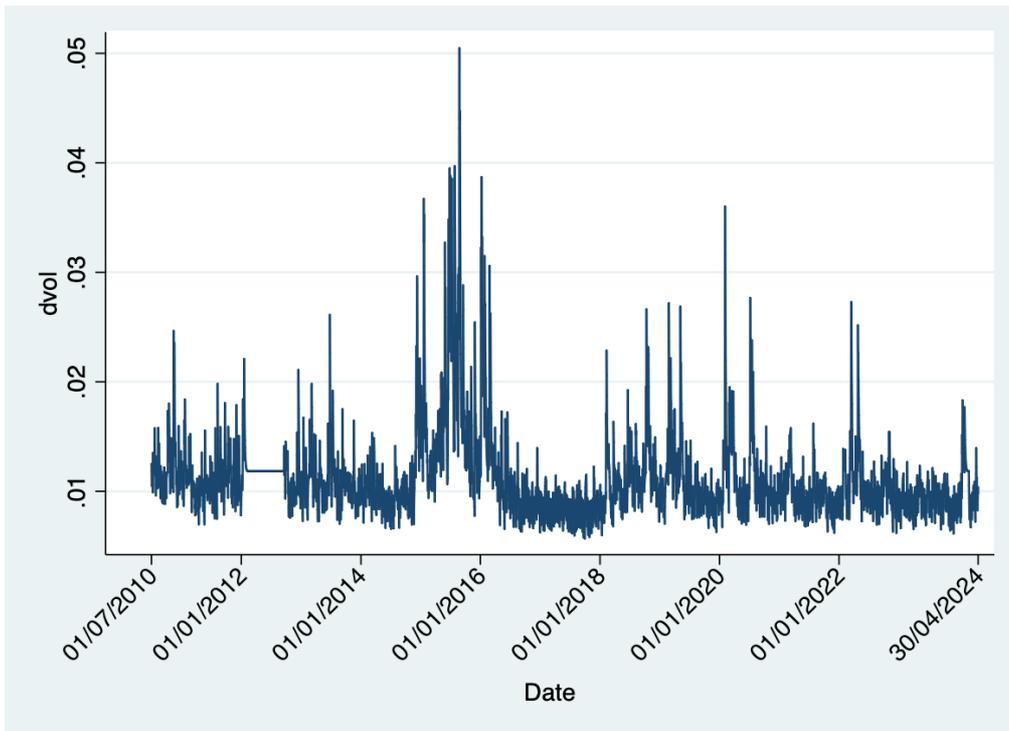

*Figure 13  Daily Stock Market Volatility*

The monthly volatility of stock market daily returns calculated by the method of Zhang Zongxin and Wang Hailiang (2013) and the volatility graphs fitted by GARCH(1,1) are roughly the same in terms of trend, and the time periods with higher volatility are mainly centered on 2015 and 2020 onwards.

4.4 Data Descriptive Statistics and Data Processing

Before proceeding to the model construction of the empirical part, descriptive statistical analysis of the variables was first carried out to get a preliminary understanding of their distribution patterns and basic characteristics such as skewness and kurtosis. The results of running with Stata are as follows:

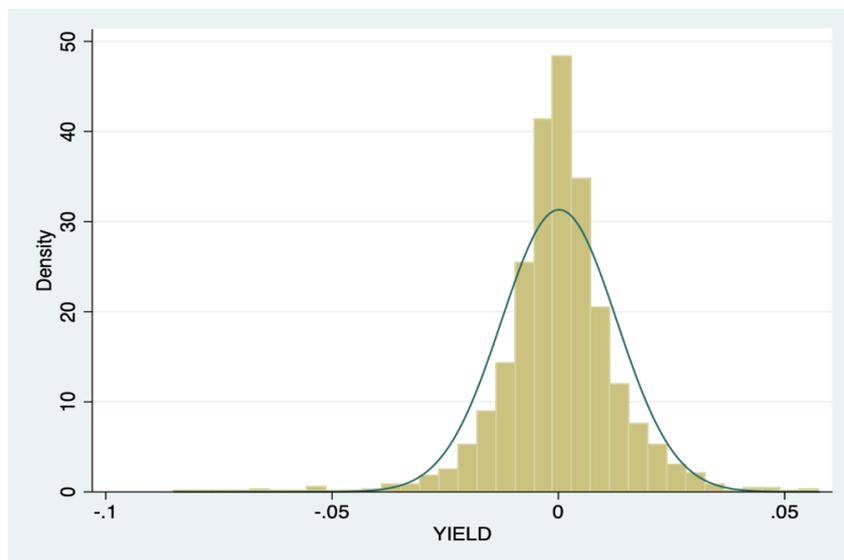

*Figure 14  Distribution plot of the variable YIELD*

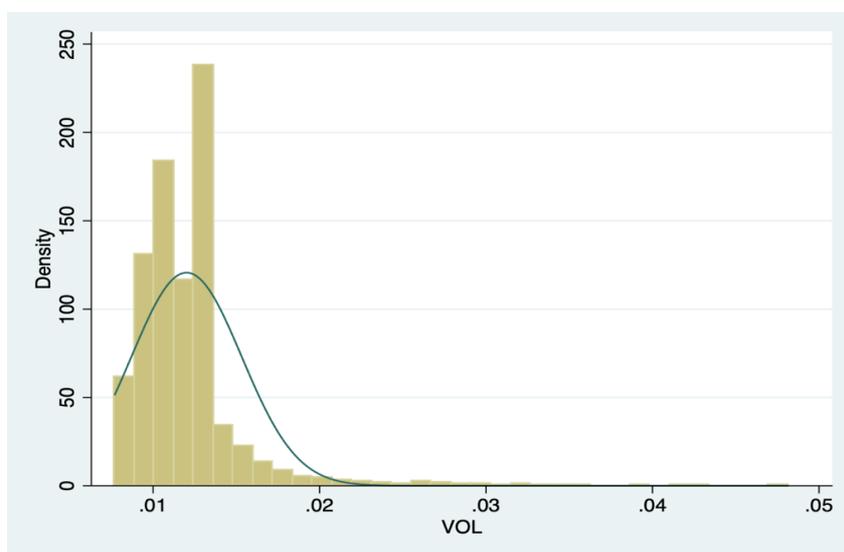

*Figure 15  Distribution plot of the variable VOL*

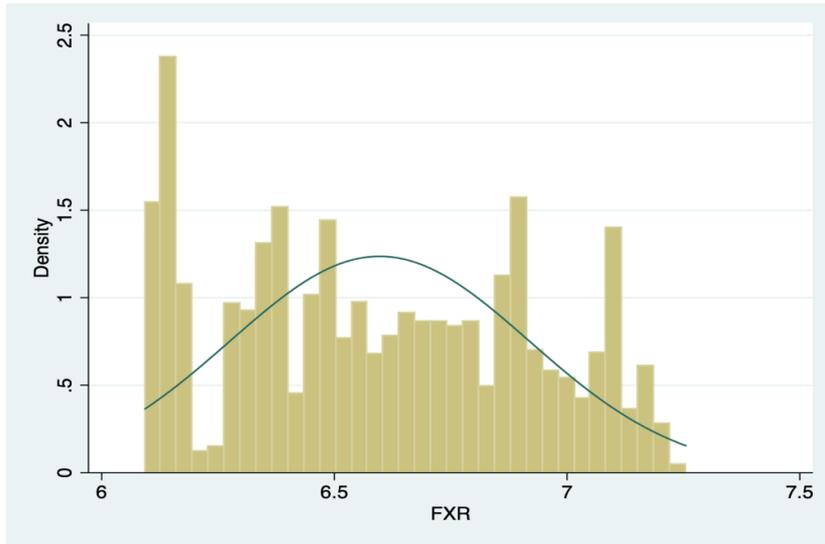

*Figure 16 Distribution plot of the variable FXR*

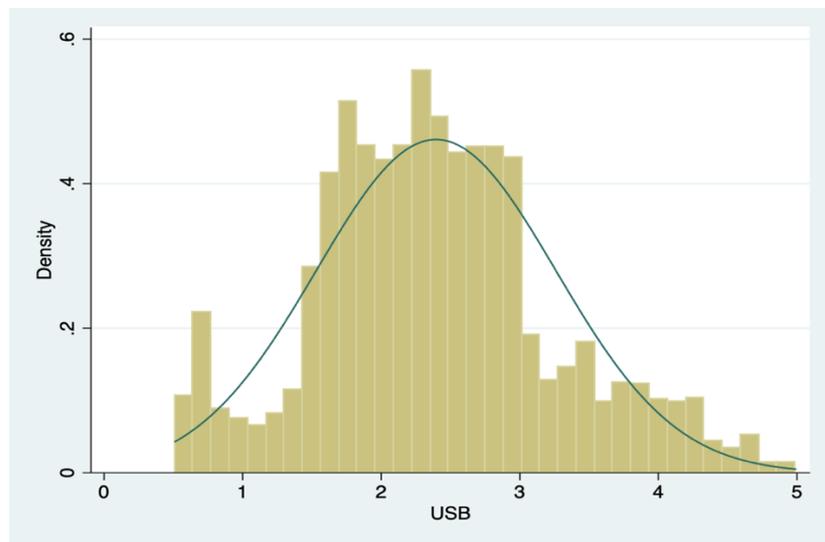

*Figure 17 Distribution plot of the variable USB*

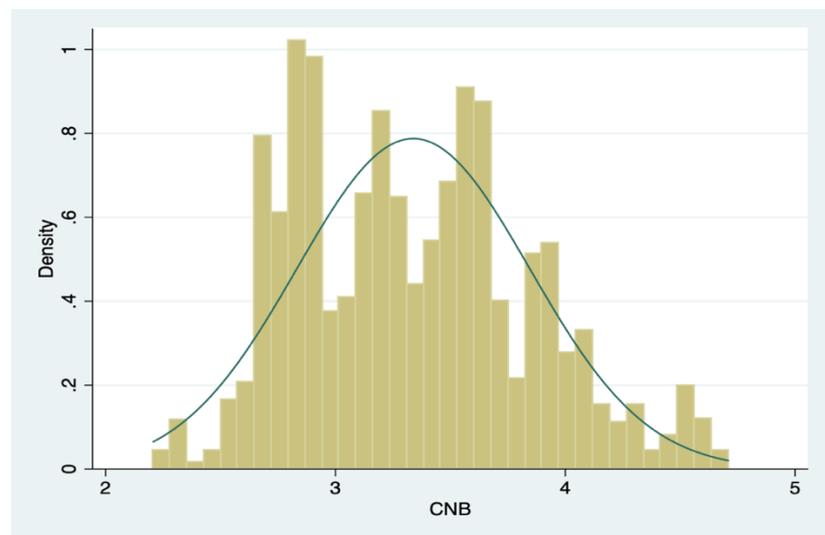

*Figure 18 Distribution plot of the variable CNB*

From the above graph, it can be initially determined that all variables do not conform to a normal distribution, but the skewness situation requires further reference to the following specific numerical indicators:

Table 2  Data Descriptive Statistics Analysis

| Variable | N | Mean | SD | Skewness | Kurtosis | Min | Max |
| --- | --- | --- | --- | --- | --- | --- | --- |
| YIELD | 3150 | 0.0087 | 0.0127 | -0.7514 | 6.0986 | -0.0849 | 0.0576 |
| VOL | 3150 | 0.0112 | 0.0039 | 3.2470 | 19.1915 | 0.0058 | 0.0505 |
| CNB | 3150 | 0.4220 | 0.5065 | 0.3645 | -0.4728 | 2.2070 | 4.7100 |
| USB | 3150 | 0.6745 | 0.8651 | 0.3157 | 0.0794 | 0.5120 | 4.9900 |
| FXR | 3150 | 0.2791 | 0.3227 | 0.1265 | -1.1486 | 6.0930 | 7.2555 |

From the above descriptive statistics table, it can be seen that except for the dependent variable Shanghai Composite Index Return (YIELD), all the other variables are right skewed. the absolute value of the skewness of daily stock return volatility (VOL) fitted by the GARCH (1,1) model is the largest, which indicates that its distribution is more skewed than the normal distribution, and the skewness of the rest of the variables is less than 3, which indicates that it is closer to the normal distribution. In terms of the peak dimension, the peaks of the return of the Shanghai Composite Index and the daily stock return volatility are relatively larger than three, indicating that their distributions are steeper than the normal distribution, which reflects that the fluctuations of stock returns have the characteristic of "sharp peaks and thick tails". In terms of the variance dimension, the variances of all variables are very small, indicating that they fluctuate narrowly around their means, but this phenomenon may be due to the small sample size itself, so the stationarity of the data series of each variable needs to be verified by more rigorous statistical tests.

5    Empirical Analysis

  5.1 Data Stationarity Test

In economics, many time series data will show a trend of large increase, large decrease or sharp fluctuation of increase or decrease over time, such as China's residents' income, consumption level, etc., which has shown a long-term trend of year-on-year increase. However, this kind of variables containing time trend characteristics in the modeling process can easily lead to model distortion, loss of reliability, the direct use of non-stationary data modeling is also prone to cause the model of the pseudo-regression problem, so the stationarity of the data needs to be tested before modeling to enhance the persuasive power. Since the sample data used in the article are also time series, the ADF test is chosen to ensure the stationarity of the data.

The ADF test is a stationarity test specifically for time series data, and the original hypothesis H0 means that there is a unit root in the sample data, i.e., the time series is not stationary. For the test results, can be judged from 2 dimensions: 1) if the ADF value corresponding to the absolute value of the T-statistic are greater than the critical value of 1%, 5% and 10% at three confidence levels, then the original hypothesis is rejected, indicating that the data are stable, through the test. 2) T-statistic corresponding to the P-value is less than the confidence level of 0.05, but also indicates that the data are stationary, through the test. On the contrary, it cannot pass the test. Therefore, in the measurement results of the ADF test, the larger the absolute value of the T statistic, the better, the smaller the P value, the better. Among them, the specific test results of each variable are shown in the table below:

*Table 3 ADF test results for each variable*

| Variable | ADF statistic | 1% Threshold | 5% Threshold | 10% Threshold | P-value | Stationarity |
|---|---|---|---|---|---|---|
| YIELD | -1.687 | -3.947 | -3.487 | -3.177 | 0.756 | Not stationary |
| YIELD(L) | -26.112 | -3.947 | -3.487 | -3.177 | 0.000 | Stationary |
| VOL | -14.235 | -3.947 | -3.487 | -3.177 | 0.000 | Stationary |
| FXR | -0.663 | -3.947 | -3.487 | -3.177 | 0.856 | Not stationary |
| FXR(L) | -7.751 | -3.947 | -3.487 | -3.177 | 0.000 | Stationary |
| CNB | -2.534 | -3.947 | -3.487 | -3.177 | 0.9963 | Not stationary |
| CNB(L) | -25.850 | -3.947 | -3.487 | -3.177 | 0.000 | Stationary |
| USB | -2.824 | -3.947 | -3.487 | -3.177 | 0.3685 | Not stationary |
| USB(L) | -27.982 | -3.947 | -3.487 | -3.177 | 0.000 | Stationary |

From the above ADF test results, it can be seen that, except for the dependent variable VOL series itself is stationary, the P-value of all the other variables itself is very large and the absolute value of T-statistic is very small, which indicates that the series is not stationary, but the P-value of the first-order differencing is at the level of the desirable value of 0 and the absolute value of T-statistic is greater than the critical value at the significance level of 1%, 5%, and 10%, which indicates that the first-order differencing of the other variables is a stationary series. The VOL series itself is stationary because it is processed similarly to the first-order difference. Therefore, the data selected in this paper are the first-order differences of VOL, FXR, CNB, and USB.

5.2 Tests for multicollinearity of variables

Since the main independent variables of the article study are exchange rate and domestic and foreign bond interest rates, and interest rates and exchange rates usually have a certain correlation, directly as an explanatory variable into the model is easy to cause the problem of multiple covariance, which leads to the phenomenon of pseudo-regression. Therefore, before the model construction of the empirical part of the first according to the correlation coefficient table for multicollinearity test, running Stata econometric software can be seen in all independent variables of the correlation coefficient table is shown below:

*Table 4  Correlation coefficients between different variables*

|     | FXR    | CNB    | USB   |
| --- | ------ | ------ | ----- |
| FXR | 1.000  | -0.645 | 0.263 |
| CNB | -0.645 | 1.000  | 0.005 |
| USB | 0.263  | 0.005  | 1.000 |

As can be seen from Table 3, the correlation coefficients of the independent variables CNB and FXR are as high as -0.645, indicating that the correlation between the variables is moderate strong, which cannot pass the test of multiple covariance, and the direct regression analysis is prone to cause pseudo-regression problems, which affects the model effect. Based on this, the article mainly adopts principal component analysis for model construction, instead of directly applying regression analysis under the least squares method. Because principal component analysis can standardize the original variable data and finally form new principal component variables with strong independence, it is a common method to solve the multicollinearity problem. Moreover, although the original independent variables in the article are not many, the lagged processing of each variable will make the study of the variables increase a lot, the use of principal component analysis can also play a certain degree of dimensionality reduction effect.

### 5.3 ARDL Model design

After the descriptive statistical analysis of the data, the basic characteristics of the variables can be initially understood. After the multicollinearity test, principal component analysis was chosen for regression analysis to eliminate the multicollinearity among the independent variables, and the construction of the model began next.

According to the model form of ARDL, the one-period lag variable of the volatility of the Shanghai Composite Index yields, the exchange rate (FXR), the Chinese government bond yield (CNB) and the U.S. Treasury bond yield (USB) and its one-period lag variable are used as the independent variables to carry out the principal component analysis, and the reason why the one-period lag of the variables is chosen is that the specific results are obtained from Stata, as shown below, the first and second principal components are denoted by F1 and F2, and so forth:

Table 5 Results of Principal Component Analysis

| Index | F1 | F2 | F3 | F4 | F5 | F6 | F7 |
|---|---|---|---|---|---|---|---|
| Eigenvalue | 3.3865 | 2.0067 | 0.9993 | 0.6012 | 0.0033 | 0.0020 | 0.0009 |
| Proportion | 0.4861 | 0.2875 | 0.1435 | 0.0819 | 0.0005 | 0.0003 | 0.0001 |
| Cumulative | 0.4861 | 0.7736 | 0.9172 | 0.9991 | 0.9996 | 0.9999 | 1.0000 |

From the above results, it can be seen that the eigenvalues of the 7 principal components in descending order are 3.3865, 2.0067, 0.9993, 0.6012, 0.0033, 0.0020, and 0.0009, respectively. Generally speaking, the eigenvalues of the principal components are greater than 1 as a criterion for the selection of principal components. Previous studies have suggested that the extracted principal components should explain at least 5-10% of the variation in the data or that the extracted principal components should cumulatively explain at least 70-80% of the variance in the data. The eigenvalues of the first two principal components are greater than 1 and the cumulative contribution rate reaches 0.7736, which means that these two principal components can explain more than 77% of the total variance. The last five principal components are omitted. The eigenvalues were examined in a steep slope plot and the graph is shown below:

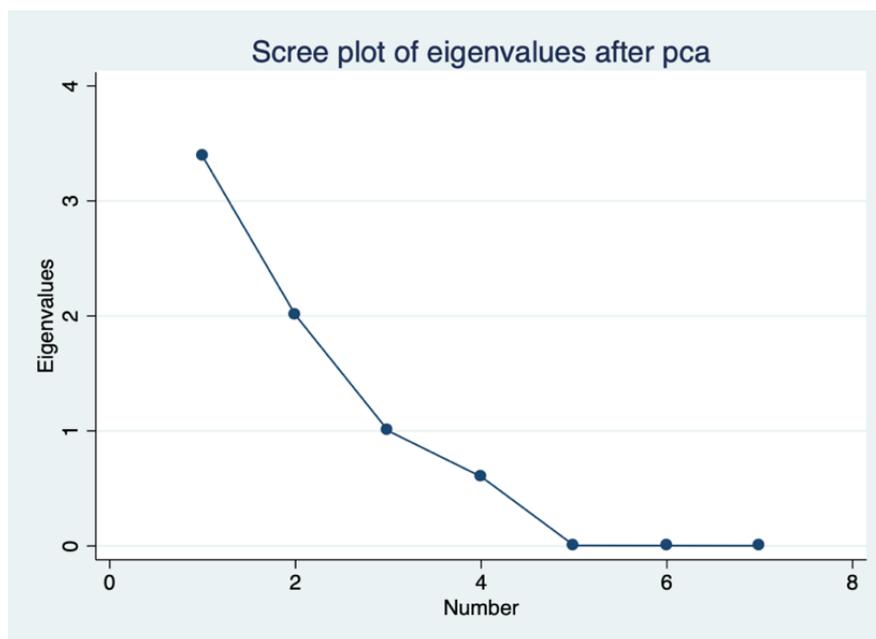

Figure 19 Scree Plot for Principal Component Analysis

It can also be seen from the above figure that the eigenvalues before sequence number 3 are larger and after that the eigenvalues are smaller, so the principal components that are retained are F1 and F2. The principal component variable score plot in Stata is shown below:

Table 6  Principal Component Loadings in STATA

| Variables | F1 | F2 |
| --- | --- | --- |
| VOL(L) | 0.00493 | 0.95855 |
| FXR | -0.27164 | -0.08663 |
| FXR(L) | -0.27132 | -0.08472 |
| CNB | 0.24877 | -0.01422 |
| CNB(L) | 0.24899 | -0.01268 |
| USB | 0.20148 | -0.03435 |
| USB(L) | 0.20131 | -0.03411 |

From the above table of coefficient loadings, it can be seen how to synthesize the new principal component variables based on the original variables after the standardization process. Thus, the eigenvectors corresponding to principal components F1 and F2 are:

$\eta_1^T = (0.00493, -0.27164, -0.27132, 0.24877, 0.24899, 0.20148, 0.20131)$

$\eta_2^T = (0.95855, -0.08663, -0.08472, -0.01422, -0.01268, -0.03435, -0.03411)$

The corresponding equations are:

$$F_1 = 0.00493 VOL(L) - 0.27164 FXR - 0.27132 FXR(L) + 0.24877 CNB \\ + 0.24899 CNB(L) + 0.20148 USB + 0.20131 USB(L)$$

$$F_2 = 0.96855 VOL(L) - 0.08663 FXR - 0.08472 FXR(L) - 0.01422 CNB \\ - 0.01268 CNB(L) - 0.03435 USB - 0.03411 USB(L)$$

From the composition of principal components $F_1$ and $F_2$, the correlation coefficient of VOL(L) in principal component $F_2$ is close to 1, while the coefficients of the remaining variables are almost 0. Therefore, principal component $F_2$ can be regarded as a proxy for the stock market intrinsic influences of the return lagged by one period of VOL(L). In principal component $F_1$, the coefficient of VOL(L) is close to 0, while the coefficients of the remaining variables FXR, CNB and USB and their lagged first-order variables are relatively more significant, which basically explains all of $F_1$, so $F_1$ can be regarded as a proxy for the stock market's extrinsic influencing factors that comprehensively reflect the exchange rate and the U.S.-China bond yields. Moreover, $F_1$ is negatively correlated with the exchange rate and positively correlated with Chinese and U.S. bond yields. Therefore, when regressing the principal components with the Shanghai Composite Index return as the dependent variable, if there is a negative correlation between the principal component $F_1$ and the stock index return, it means that there is a positive correlation between the return and the exchange rate proxies, FXR and FXR(L), and a positive correlation with the interest rate proxy variable, CNB.

The negative correlation between principal component $F_1$ and stock index returns indicates that returns are positively correlated with the exchange rate proxies FXR and FXR(L), and negatively correlated with the interest rate proxies CNB, CNB(L), and USB, USB(L). Summarizing the above analysis and the aforementioned theories, under different theoretical foundations and transmission mechanisms, there is not a unique and definite correlation between the RMB exchange rate and interest rate and China's stock market, but an uncertain relationship that can be positive or negative. Previous studies also show that the impact of interest rate and exchange rate on the return of China's stock index varies in different periods of the capital market. However, usually when our stock market runs stationarily and performs well, there is a negative correlation between the U.S. Treasury bond yield and our stock market, so if it is to be consistent with the theory, the correlation between $F_1$ and the dependent variable should be negative. In addition, considering the time lag effect of the capital market and China's A-share market perennial sideways oscillations, ups and downs, the specific impact of the historical performance of stock prices on the stock market cannot be accurately measured, but the author believes that the negative correlation may be more accurate, because in the short term, the ups and downs is the stock market's normal, the stock price has been upward in the bull market and down all the way in the bear market belong to the more special period.

Thus, in the equation with the performance of the Shanghai Composite Index as the dependent variable, this paper proposes the research hypotheses:

$F_1$ proxies for the extrinsic stock market influences that comprehensively reflect the exchange rate (FXR), the Chinese and U.S. bond yields (CNB, USB), and their first order lagged variables;

$F_2$ proxies for the intrinsic stock market influences that are the lagged one-period VOL(L) of the yield itself. It is also assumed that the interrelationships between these two independent variables and the dependent variable are both negative.

After selecting the principal components, the correlation between them is basically eliminated, and then regressing them on the dependent variable can effectively prevent the pseudo-regression phenomenon and improve the accuracy of the model. However, the regression of principal components can only ensure the elimination of multicollinearity of variables, and other requirements such as significance test and heteroskedasticity test may not be realized, so the extracted principal component regression equations need to be tested more.

In the modeling process, first try a simple multivariate linear model, although the article's principal components only choose 2, but the actual variables represented by 7, set:

$$VOL = C_0 + C_1 * F_1 + C_2 * F_2 + \varepsilon$$

The following results are obtained by applying STATA:

*Table 7 Multivariate regression result*

| Explanatory Variable | Coefficient | Standard Error | T-value | P-value |
|---|---|---|---|---|
| C | 0.0120105 | 0.0000579 | 207.46 | 0.000 |
| F1 | 0.002741 | 0.0000314 | -8.73 | 0.000 |
| F2 | 0.002597 | 0.0000408 | -6.36 | 0.002 |

Result can be obtained as:

$$VOL = 0.002741 F_1 + 0.002597 F_2 + 0.0120105$$

From the above results, it can be seen that the P-values corresponding to F1 and F2 are small, indicating that the model can pass the significance test. And the overall P-value under the F-statistic is also less than 0.05, indicating that the significance between the variables as a whole is strong. In addition, the residual sum of squares is also small, indicating that the model is accurate. The results show that the statistical indicators are more reasonable, which all indicate that the multivariate linear equations mentioned above are satisfied between the principal components and the dependent variables, and it can be initially judged that the model is effective, but ultimately, it needs to pass a more rigorous special test.

### 5.4 ARDL Model testing

In order to pursue academic rigor, the model is tested with the help of STATA software as follows, which mainly includes ADF test of variables, cointegration test, Granger causality test, and heteroskedasticity test to further rule out the phenomena of multicollinearity and pseudo-regression. Among them, the ADF test has been described in the previous section (5.2), so the cointegration test, Granger causality test and heteroskedasticity test are put into this subsection.

#### 5.4.1 Co-integration test

The analysis of ADF in section 5.2 shows that some independent variables are stationary while others are not, so the Granger causality test cannot be conducted directly. However, the variables are all stationary after the first-order differencing, which satisfies the precondition for the co-integration test, and the co-integration test can also show that there is a long-term stable equilibrium relationship between the explanatory variables and the explained variables. There are EG two-step method and Johansen test for co-integration test, but the first method is mainly based on the co-integration between two variables, and it is a test of model residual term. The Johansen test, on the other hand, is based on multivariate variables and is a test of the model correlation coefficients, which usually has a higher level of confidence. Since the article studies the correlation between multivariate variables, Johansen's test is chosen to verify the co-integration relationship, and the specific running results in STATA are as follows:

*Table 8  Co-integration test result*

| Co-integration assumption | Eigenvalue | Trace statistic | Trace P-value | Max trace statistic | Max trace P-value |
|---|---|---|---|---|---|
| No cointegrating relationship | 0.1525 | 221.4897 | 0.0001 | 137.3498 | 0.0001 |
| One cointegrating relationship | 0.0964 | 84.1398 | 0.0000 | 84.1218 | 0.0000 |
| Two cointegrating relationships | 2.18E-05 | 0.0181 | 0.8929 | 0.0181 | 0.8929 |

From the above results, it can be seen that both the trace statistic test and the maximum eigenvalue statistic test, under the hypotheses of no co-integration and up to one co-integration relationship, their corresponding P-values are less than the 1% significance level, so the original hypothesis is rejected. Similarly, under the hypothesis that there are 2 co-integration relationships, their P-values are greater than the significance level. Therefore, the co-integration test is passed, indicating that the exogenous proxy variable F1 of exchange rate, Chinese and US bond yields and the endogenous proxy variable F2 of stock index's own historical performance can indeed affect China's stock market and there is a stable long-term equilibrium relationship.

### 5.4.2  Granger causality test

Although the Granger causality test cannot be performed directly due to the stationarity problem of the data, the Granger causality test can be performed successfully after passing the cointegration test. It should be clear that the Granger causality test does not indicate that there is a causal relationship between the variables, but reflects the essence of the temporal order in which the variables change, i.e., if A is the Granger cause of B, it means that A changes before B. The results of the model using the STATA econometric software are summarized as follows:

*Table 9 Granger causality test result*

| Hypothesis | Sample number | F-statistic | P-value |
| --- | --- | --- | --- |
| F1 is not a Granger reasons for VOL | 1347 | 2.3602 | 0.1240 |
| VOL is not a Granger reasons for F1 |  | 1.5173 | 0.2199 |
| F2 is not a Granger reasons for VOL | 1347 | 5.3123 | 0.0051 |
| VOL is not a Granger reasons for F2 |  | 1.1578 | 0.6911 |

As a result, no matter at 1%, 5% or 10% level of significance, according to the P-value, it can be obtained that: the original hypothesis that the proxy variable F2 is not the Granger cause of the VOL of last day's stock index return needs to be rejected, i.e., F2 is the Granger cause of the return of last day's stock index, which indicates that the change of F2 comes before. Since F2 is a proxy variable for the previous day's stock index itself, and the result of the Granger test is that the change of F2 occurs first, it further verifies the feasibility of using F2 as a proxy variable for the return of the previous day's stock index. All the rest of the original hypotheses are accepted due to the large P-value. However, according to the principle and practical significance of the test, it does not indicate that there is no causal relationship between the independent variables and the dependent variable, but only indicates that due to the strong random volatility of the capital market, the sequence of changes between the variables can not be accurately measured, and the Granger test is only one of the tests often conducted for time series data.

### 5.4.3 Heteroscedasticity test

In least squares analysis, it is necessary that the residuals of the regression equation conform to a heteroskedastic distribution, i.e., the variance of the residuals is fixed for all residuals and does not change with the change in the reference value. In STATA, the White's test can be used to verify the heteroskedasticity, which is based on the assumption that the model meets the requirement of heteroskedasticity, and then estimate the regression equation, and then test the estimated residuals to see if they really meet the assumption of heteroskedasticity. White's heteroskedasticity test is performed and the results of the analysis in STATA are shown below:

*Table 10 Heteroscedasticity test result*

| F-statistic | 0.9487 | P-value | 0.1784 |
|---|---|---|---|
| $nR^2$ | 18.0680 | Prob > chi2 | 0.3882 |

As can be seen from the above figure, the model corresponds to an F-statistic of 0.9487 and the p-value under the F-statistic is 0.1784. At 5% significance level, the original hypothesis of the existence of heteroskedasticity is accepted since the p-value is greater than 0.05, i.e., the model does not suffer from heteroskedasticity, which means that the obtained estimates of each parameter are valid and do not need to be further corrected.

5.5 Analysis of empirical results

Up to this point, the model has solved the problem of multicollinearity through principal component regression, and through the ADF unit root test, it shows that the first-order differences of the variables are stationary and have the conditions for cointegration test. And the cointegration test of the correlation coefficients shows that there is a long-run equilibrium relationship between the variables. In addition, Granger causality analysis was also carried out, and heteroskedasticity test of the residuals indicated that the residuals conformed to the expectation of homoskedasticity, which further ruled out the possibility of pseudo-regression, and it can be seen that the design of the model is more reasonable.

In the following, the regression results are analyzed and interpreted. First of all, according to the regression equation, it can be seen that the negative correlation between the dependent variable and variables F1 and F2 are established, which is in line with the assumptions and expectations of the study. Among them, the correlation coefficients of F1 and F2 are larger, indicating that under the combined effect of the exchange rate and the yields of Chinese and U.S. bonds, superimposed on the historical performance of the stock index itself, it can have a more significant impact on China's Shanghai Composite Index. When China's economy runs stationarily and the stock market performs well, there is a negative correlation between variable F1 and the return of the Shanghai Composite Index, while there is a negative correlation between F1 and the USD-RMB exchange rate FXR and its first-order lagged variables, and a positive correlation with the yields of Chinese and U.S. bonds (CNB, USB) and their first-order lagged variables. Therefore, the model finally shows that the return of China's Shanghai Composite Index is positively correlated with the exchange rate factor and negatively correlated with the Chinese and U.S. bond yields, and of course, this relationship also applies to the one-period lag of each variable.

Specifically, as far as the relationship between exchange rate and stock price yield is concerned, the optimal lag between exchange rate and stock price is one period, following the competitiveness effect, i.e., under the direct markup method, when the exchange rate between the previous day and the current day rises, the RMB depreciates, making the domestic goods more competitive, increasing exports and decreasing imports, which leads to the trade surplus, which, on the one hand, makes the foreign exchange reserves increase, and on the other hand Domestic GDP rises, which ultimately provide strong support for the economic fundamentals of China's A-share market, making share prices rise and yields rise. The study shows that the positive relationship between exchange rate and stock index yields may be due to the phenomenon of unusually sharp volatility of the stock market with large upward and downward movements during the sample period of the study.

In terms of the relationship between Chinese treasury bond yields and stock index yields is concerned, the optimal lag period is also one period, following the total social supply and demand effect and the asset portfolio effect, i.e., when the Chinese treasury bond yields rise on the previous day and the current day, on the one hand, it makes the investment yields on equities relatively decline, the funds entering the stock market decrease, and the demand for equities falls. On the other hand, as an important reference index of market interest rates, the rise of treasury bond yields will also lead to the rise of other market interest rates, which makes the investment and consumption decrease, the total social demand decline, the market environment deteriorates, the business performance of enterprises is dismal, which in turn inhibits the investment demand for stocks, and ultimately makes the stock price go down and the yield fall.

In terms of the relationship between the U.S. Treasury yields and the Shanghai Composite Index, the negative correlation is also in line with the theory, and the best lag shown in the model is one period, which indicates that when the U.S. Treasury yields of the same day and the previous day increase, investors will increase the demand for U.S. Treasuries due to the tendency for funds to be profit-making, which, on the one hand, makes the market less liquid, and the funds entering the A-share market of China are reduced accordingly. On the other hand, the market demand for US dollars increases, the US dollar appreciates, the RMB depreciates, and the central bank sells US dollars and buys RMB in the foreign exchange market. Ultimately, this leads to a decrease in the base currency of the RMB, which in turn leads to a decrease in stock prices and a decrease in yields.

In addition, as far as the effect of variable F2 is concerned, it enters the model with optimal lags of order 1 and order 2, respectively, and F2 itself is a proxy variable for the one-period lag of the stock index, which suggests that the stock market's optimal lags for itself are order 2 and order 3. In addition, the negative correlation suggests that if the stock index closed higher on the previous day and the day before that, it tends to make the stock market fall on the same day, which may be due to the market expectation that when the stock price goes up in the previous period, investors expect that the stock price will eventually go down, which leads to a decrease in the demand for stocks and a decrease in the rate of return, which is also in line with the long term phenomenon of the stock market that has a long history of alternating ups and downs, and constant oscillations. Finally, from the point of view of the time trend term, the performance of China's stock index is also affected by the time factor, which is mainly manifested in the phenomenon of random fluctuations with the advancement of time. Overall, the study shows that the exchange rate, Chinese and U.S. bond yields and the historical performance of the stock index itself are indeed the factors that can have a more significant impact on China's stock market, and the exchange rate, Chinese and U.S. Treasury yields correspond to the best lag of 1 period, while the best lag of the stock index itself is 2 and 3 periods, and the performance of the stock index has a clear time trend characteristics.

## 6  Conclusion and Suggestions

With the stationary addition of RMB to the SDR basket of currencies in 2015, the implementation of Shanghai-Lunan Tong in 2016, the attempt of SSE Composite Index to be included in the MSCI index of emerging markets, the impact of Fed's interest rate hike and Brexit expectations on China's A-share market, and the market concerns about the decline of foreign exchange reserves and capital outflows also pressurized the A-share market, there are multiple indications of the increasingly closer ties between China's capital market and the international market. Based on this, this paper investigates the impact of the spot exchange rate of RMB against USD and domestic and foreign bond yields on China's Shanghai Composite Index, which not only reflects the impact of domestic and foreign factors on China's stock market, but also reflects the linkage of the foreign exchange market, bond market and stock market.

The interrelationships and transmission mechanisms between exchange rates, bond market interest rates and stock prices are theoretically studied separately. Among them, the transmission of exchange rate to stock price is analyzed mainly from three aspects: competitiveness effect, cheap import effect and inflation effect. The transmission of interest rate to stock price is analyzed mainly from three aspects: total social supply and demand effect, social wealth effect and asset portfolio effect. Empirically, the multicollinearity between variables is resolved by using principal component analysis, which is used as a proxy variable for the respective variables according to the composition of the principal components. In order to reflect the time lag effect, the ARDL model was chosen to lag the variables, and passed the ADF unit root test, cointegration test, white noise test and heteroskedasticity test, which indicates that there is a long-run equilibrium relationship between the variables.

The final model shows that the return of China's Shanghai Composite Index is positively correlated with the exchange rate variable, while it is negatively correlated with the treasury bond yields of China and the United States and its own lagged variables, which proves that China's stock market is indeed affected by domestic and foreign factors together. Among them, the positive correlation between the exchange rate and stock index returns may be due to the phenomenon of unusually sharp volatility of the stock market with big rises and big falls during the study sample period. Overall, the overall exogenous influences of the RMB exchange rate and Chinese and U.S. bond yields can have a more significant impact on China's Shanghai Composite Index returns, as can the historical factors of the stock index itself, as well as the time factor. Combined with the theory, when China's stock market performs well, the positive correlation between the exchange rate and the stock market is mainly transmitted to the stock market through the competitiveness effect. And as far as the impact of China's treasury yields on the stock market is concerned, the results follow the total social supply and demand effect and the asset portfolio substitution effect, with a negative correlation between the stock market. The model shows that the exchange rate, China, the United States Treasury bond yields of the best lag are 1 period, while the stock index itself is the best lag of 2 and 3 periods, respectively, if the day before yesterday and the day before the big day before the stock index closes up, then it is easy to make the stock market fell on the same day, which may be due to the market expectations, the stock market is expected to go down when it rises, and when it falls, it is expected to go up.

As the pace of economic globalization deepens in the future, the influence of foreign factors on China's A-share market is expected to continue to deepen. Therefore, in order to maintain the stationary operation of China's stock market, we should formulate a mature and perfect economic system, unclog the interconnection channels among the stock, foreign exchange and bond markets, and resolutely implement all-round financial reforms to strengthen the marketization and internationalization of the A-share market. In addition, it should also hit the current pain points of the stock market, strengthen capital supervision, severely punish illegal fund-raising, and purify the competitive environment of the market.

Compared with developed countries such as the U.S., China's capital market started late but developed fast. the establishment of the Shanghai Stock Exchange in December 1990 marked the beginning of the development of China's stock market, and the Chinese stock market has expanded rapidly in just 20 years. As of November 17, 2016, the total market value of China's Shanghai and Shenzhen stock markets reached about 53 trillion yuan, of which the circulating market value is nearly 41 trillion yuan, behind the huge volume also inevitably contains many contradictions and risks. As stocks are a kind of virtual capital, coupled with excessive speculative behavior in the securities market, stock prices often experience sudden and huge fluctuations, causing huge losses to investors and disrupting the market order. In addition, China has yet to establish a set of sound stock market support system and regulatory system, the lack of integrity of listed companies and incomplete disclosure of information leads to information asymmetry, lagging effect of policies, and excessive speculative atmosphere and so on. In view of the phenomenon of frequent sharp fluctuations in China's stock market and the pain points of the market, such as illegal fund-raising, as well as the conclusions drawn from the empirical model, the author puts forward the following three recommendations.

6.1 Accelerating domestic stock marketization and internationalization

According to the conclusion of the empirical analysis, it can be seen that the performance of China's stock index is jointly influenced by the bond market, the foreign exchange market and foreign markets, and the cointegration test also illustrates the existence of a stable long-term equilibrium relationship between the exchange rate, domestic and foreign interest rates and the A-share. Therefore, to enhance the return and stability of China's stock market, it is necessary to accelerate the pace of marketization and internationalization of A-shares, implement multi-level and all-round reforms so that the market can fully reflect the interaction of the stock, foreign exchange, and bond markets, and at the same time pay more attention to the changes in overseas markets. Specifically, including the simultaneous marketization of interest rates and exchange rates; gradually allow foreign investors to foreign exchange hedging of RMB onshore market exposure, rather than just the offshore market; rhythmic relaxation of capital project control; reduce the investment thresholds and restrictions on foreign investors to enter China's stock market requirements, and then stationary the flow of foreign capital into China's stock market channels, the introduction of diversified, internationalized investment subjects and so on.

Nowadays, the status of China's market economy has been basically established, and in recent years, the pace of economic and financial marketization has accelerated significantly. the reform of the RMB exchange rate system in August 2015, and the opening of the Shenzhen-Hong Kong Stock Connect in November 2016, etc. are all the products of the implementation of market-oriented reforms in China's capital market. Secondly, corporate IPOs have completed the transition from the filing system to the registration system, and the phenomenon of companies queuing up to go public has eased, which is also a market-oriented reform initiative to better utilize the stock market's financing function and meet the capital needs of enterprises. In addition, the melting mechanism was implemented at the beginning of last year in order to curb the sharp fall of stock prices, although the final results of the practice showed that the melting mechanism is not adapted to China, and its life span in the A-share market is only two days, because of its magnetic attraction effect, which accelerates stock market rises or plummets at the threshold, thus triggering the meltdown. But this is also in the market-oriented reform of the road of continuous experimentation, trial and error process. Reform needs to pay a price, but only reform can solve the contradiction between the rapid development of China's stock market and the lack of internal systems, in order to support and nurture the healthy and sustainable development of China's stock market.

Compared to the pace of marketization, the internationalization of A-shares has yet to speed up. Although the RMB has been included in the "basket" of currencies, the attractiveness of China's stock market to international investors is still far less than that of developed capitalist countries such as the United States, partly because of China's tighter capital controls and the very limited and quota-constrained channels available for foreign investors to invest in A-shares. Partly because of China's tight capital controls, foreign investors have very limited channels to invest in A-shares, and there are also quota restrictions. However, standing in the wind of economic globalization, China's stock market should continue to strengthen the links and exchanges with the international market, loosely or tensely liberalize the capital control, follow the law of free flow of capital, enhance the level of internationalization, and pay close attention to the Fed's interest rate hike, the United Kingdom's exit from the European Union and other international events that may affect China's stock market, and if necessary, should also take corresponding countermeasures.

6.2 Unblocking equity, currency and bond market linkages

The model shows that the stock index was not only significantly affected by the performance of the stock index in the previous two days, but also affected by the bond market and foreign exchange market at the same time. Good economic fundamentals is to ensure the balanced and stable development of China's stock, foreign exchange, bond and other capital sub-markets, to maintain the fundamentals of the good must be prompted by China's economic growth to achieve progress in a stable manner. As we all know, the stock market is a reflection of a country's economic situation of the barometer, the economy is good or bad first from the stock market on the direct performance of the stock market, the economy is good when the stock market can usually enter the upward channel, the economy is poor when the stock market is followed by atrophy. The mature and perfect economic system is to promote balanced economic development of the basic guarantee. Perfect economic system not only includes the stock market related supporting system, should also include the bond market, the exchange market related system, so as to effectively promote the circulation of funds in the three cities and transmission, accelerate the speed of currency circulation and economic operation efficiency.

As can be seen from the model, the exchange rate and the stock market is a positive correlation between the interest rate and the stock market is a negative correlation between the stock market, which also means that when the stock market fell more than, the central bank can be adjusted through the foreign exchange market to raise the RMB exchange rate and the bond market to reduce the market interest rate, so that the stock market to play a certain degree of help to the role of the upward movement. Of course, investors can also pay full attention to the market performance of interest rates and exchange rates to determine whether it is the right time to invest in stocks. Therefore, unclogging the connectivity channels among the three markets is also conducive to enriching the national macro-control tools.

In addition, a mature economic system should strengthen the trading function of the stock market. From the perspective of economic globalization, the competition among financial markets in various countries is becoming increasingly fierce, and the competition in the stock market is mainly manifested in the competition among various exchanges. Usually, the more comprehensive and powerful the functions of a stock market, the more attractive it is to capital, and the more competitive it is accordingly. The realization of the stock market function must be based on a reasonable stock trading system, because the stock trading system can not only convert potential investor demand into actual transactions, but also by affecting the transaction costs of investors and then flexibly adjust the market liquidity, increase market transparency, and further affect the pricing of the stock and price fluctuations.

Currently, China's Shanghai and Shenzhen stock exchanges, treasury bonds, corporate bonds, convertible bonds and warrants is the implementation of the "T + 0" trading system; A shares, the fund is the implementation of the "T + 1" trading system. This kind of flexible and efficient trading system was gradually conceived through continuous reform and market verification, which can, to a certain extent, make the rigid and inefficient trading mechanism of China's market situation initially improved.

However, the stock exchange system should gradually move towards the goal of better function and more reasonable pricing mechanism. This requires paying closer attention to the relationship between the market system and pricing behavior, focusing on the impact of information asymmetry on the behavior of market participants, and solving the problems of free-riding, moral hazard and adverse selection resulting from market failure. And then comprehensively study the design of the trading system, set a complete set of mature trading process norms, and ultimately through reliable price discovery, extensive price dissemination and effective price risk avoidance to make China's stock market more efficient and stable operation.

Secondly, the author believes that in order to purify the capital market of good and bad competition environment, can flexibly introduce the delisting system, will be long years of low performance, benefits are not outstanding, no industrial support or only rely on the speculation of the shell resources to guide the delisting, so as to make room for more outstanding enterprises to go public, reduce the threshold for listing. On the one hand, not only can incentivize listed companies to strive to maintain good profitability, so that the enterprises remain in the market are blue chip stocks with excellent performance, reduce the demon stock fan line, through the market selection to provide investors with good investment targets. On the other hand, it can also combat the enthusiasm of the market speculation, advocate the development of industry, and guide the capital off the virtual into the real.

### 6.3 Strengthening capital market regulation

At present, China's capital market regulation is still characterized by the lack of effectiveness and authority of the regulatory bodies, insufficient innovation in capital market regulation theory, lack of self-regulation, and inadequate legal construction of market regulation. As a result, leading to the financial institutions to play the policy of rubbing the ball and and illegal fund-raising phenomenon is serious, high leverage prying huge sums of money to engage in high-risk transactions and even violations of market manipulation problems are commonplace. To address the pain points of the stock market, the author believes that regulation can be strengthened in the following three aspects:

I. With the help of the power of the Internet, to build a comprehensive system of risk warning indicators for capital market regulation, to quickly find abnormal reflections of relevant macroeconomic indicators and timely regulation, and to strengthen the real-time monitoring of large trading funds and trading venues, thereby preventing or avoiding the occurrence of financial risks.

II. Taking the protection of the legitimate rights and interests of public investors as the main objective of regulation, promoting the diversification of regulatory means, strengthening regulatory legislation and the construction of industry self-regulatory organizations, clarifying the main functions and rights of each regulatory body, and conducting in-depth research on regulatory blind spots and blank areas.

III. Establishing specialized agencies to comprehensively grasp and evaluate the integrity of market entities, increasing the disciplinary efforts against non-compliant entities in administrative licensing and regulatory enforcement, conducting resolute, timely and public investigations and punishments for trading in power and money, illegal fund-raising and market manipulation in violation of the law, and setting up a unified database of information on the integrity of the capital market to implement a unified management of the integrity records of market entities.

The ability to regulate finance and control risks has become a very important national soft power. At present, China's financial market supervision is mainly carried out by the Banking Regulatory Commission (CBRC), the Insurance Regulatory Commission (IRC) and the Securities Regulatory Commission (SRC). However, there are many contradictions and frictions in the separate regulation, which makes the overall supervision inefficient and is not conducive to the synergistic development of the capital submarkets. In particular, standing on the windfall of big data and Internet+, cross-border financial derivatives and innovative products are emerging, and insurers are frequently using their strong financial strength to get involved in the capital market and raise their stakes in listed companies. As a result, since 2015, the market has gradually appeared the voice of joint supervision, and the author is also looking forward to the realization of the merger of the three chambers under the unified leadership of the central bank. After all, in an era when financial innovation is exploding, the launch of a variety of innovative businesses and financial products may simultaneously involve the integration of the stock market, bond market and foreign exchange market, and the lack of unified supervision will lead to numerous gray areas, which will disrupt the market order.

Of course, effective regulation does not only rely on the government to make every effort and industry self-regulatory behavior, investors themselves should also rationally deal with the changes in the stock market, strengthen their own quality training, precipitate the market contract spirit and self-responsibility. In a standardized, fair and sound market, truly responsible for their own behavior and profit and loss. Finally, in order to move closer to a mature capital market, China's A-share market should cultivate more institutional investors and more professional financial investment talents, so as to be able to grasp the operating rules of the unpredictable capital market in a timely manner, to reduce severe market shocks and to strengthen the risk prevention. Although the mature capital market is not a large number of institutional investors as a necessary condition, but the current Chinese A stock market, too many retail investors, speculative behavior is too heavy, information asymmetry and lack of investment expertise phenomenon is more obvious. So the development of institutional investors and professional investment talent becomes extremely necessary. Can also be expanded through the proportion of direct financing, reduce the cost of corporate fund-raising, build an open and transparent information disclosure platform for listed companies, strengthen the information disclosure system in the capital market, improve the information disclosure of dynamic regulatory mechanisms, the implementation of the capital market accounting and auditing system, standardize the organizational structure of listed companies, strengthen the use of funds raised by the enterprise and supervision of the direction of the investment, and thus strengthen the listed companies' profitability and risk-resistant ability. A series of initiatives to promote the long-term stable and efficient development of the capital market.